\documentclass[
reprint,
superscriptaddress,
nofootinbib,
prb,
amsmath,amssymb,
aps,
]{revtex4-1}
\usepackage[english]{babel}
\usepackage[utf8]{inputenc}
\usepackage{amsmath,amssymb,amsfonts}
\usepackage{color}
\usepackage{graphicx}
\usepackage[colorlinks=true,linkcolor=blue,urlcolor=blue,citecolor=blue]{hyperref}

\begin{document}

\title{Conductance of quantum spin Hall edge states from first principles: the critical role of magnetic impurities and inter-edge scattering}

\author{Luca Vannucci}
\email{lucav@fysik.dtu.dk}
\affiliation{CAMD, Department of Physics, Technical University of Denmark, 2800 Kgs. Lyngby, Denmark}

\author{Thomas Olsen}
\affiliation{CAMD, Department of Physics, Technical University of Denmark, 2800 Kgs. Lyngby, Denmark}

\author{Kristian S. Thygesen}
\affiliation{CAMD, Department of Physics, Technical University of Denmark, 2800 Kgs. Lyngby, Denmark}
\affiliation{Center for Nanostructured Graphene (CNG), Technical University of Denmark, 2800 Kgs. Lyngby, Denmark}

\date{\today}

\begin{abstract}

The outstanding transport properties expected at the edge of two-dimensional time-reversal invariant topological insulators have proven to be challenging to realize experimentally, and have so far only been demonstrated in very short devices.
In search for an explanation to this puzzling observation, we here report a full first-principles calculation of topologically protected transport at the edge of novel quantum spin Hall insulators --- specifically, Bismuth and Antimony halides --- based on the non-equilibrium Green's functions formalism.
Our calculations unravel two different scattering mechanisms that may affect two-dimensional topological insulators, namely time-reversal symmetry breaking at vacancy defects and inter-edge scattering mediated by multiple co-operating impurities, possibly non-magnetic.
We discuss their drastic consequences for typical non-local transport measurements as well as strategies to mitigate their negative impact.
Finally, we provide an instructive comparison of the transport properties of topologically protected edge states to those of the trivial edge states in MoS$_2$ ribbons.
Although we focus on a few specific cases (in terms of materials and defect types) our results should be representative for the general case and thus have significance beyond the systems studied here.

\end{abstract}

\maketitle

\section{Introduction}
\label{sec:intro}

In a pair of groundbreaking papers published in 2005, Charles Kane and Eugene Mele argued that graphene becomes a two-dimensional topological insulator (2D TI) at sufficiently low temperature \cite{Kane05_1,Kane05_2}.
In other words, a flake of graphene cooled down to cryogenic temperature will essentially behave as a band insulator everywhere except on its boundaries, where special metallic edge states will appear.
Such states are immune to scattering against impurities or disorder, and therefore realize a perfect dissipationless conductor with great potential for future technological applications.
The reason behind this striking robustness is rooted in the mathematical concept of topology, hence the name \emph{topological insulators} \cite{Haldane17_Nobel}.

Unfortunately, the band gap opened by spin-orbit coupling (SOC) in graphene is actually too small to give rise to any measurable effect \cite{Yao07}.
To realize the first 2D TI --- or quantum spin Hall insulator (QSHI) --- it was therefore necessary to resort to quantum-well heterostructures, which indeed showed the much anticipated signatures of topologically protected transport in non-local multi-terminal measurements \cite{Konig07,Knez11}.
With the rising awareness that the realm of monolayer 2D materials is actually much larger than initially thought, several new QSHIs have been reported in other 2D monolayers than graphene in recent years \cite{Reis17,Wu18_WTe2_exp,Kandrai19,Cucchi19}.

It is, however, not fully understood to what extent the ideas of topological protection can materialize into the next generation of electronic devices, due to some inconsistency between theory and experiments.
Time-reversal symmetry (TRS) forbids electron backscattering on the edge of 2D TIs, since counter-propagating modes have opposite spin polarization --- they are therefore termed \emph{helical edge states}.
The defining feature of such a pair of protected states is a well defined quantized conductance plateau at $G_0 = 2 e^2/h$, which is, however, difficult to attain in the lab. The few successful attempts are all limited to very low temperature ($\sim 1$K in quantum-well heterostructures \cite{Konig07,Knez11}) or very short channels ($\sim 100$nm for the case of monolayer WTe$_2$ \cite{Wu18_WTe2_exp}).
Attempts to understand this discrepancy at the model level have focused on many diverse backscattering mechanisms driven by electron-electron interactions, charge puddles, embedded nuclear spins, coupling to phonons and electromagnetic noise \cite{Wu06,Xu06,Maciejko09,Strom10,Tanaka11,Budich12,Schmidt12,Vayrynen13,Hsu17_nuclear-spins,Hsu18_nuclear-spins,Groenendijk18,Vayrynen18,Novelli19}.
Nonetheless, the question is still much debated and deserves a careful analysis from a different and more realistic point of view.

Here we report for the first time a full first-principles study of topologically protected transport at the edge of novel QSHIs.
We use newly developed computational 2D materials databases \cite{Mounet18,Haastrup18}, containing existing structures as well as hitherto unknown monolayers, to identify a family of large-gap QSHIs ideally suited for the current study.
We then explore the electronic band structure of such candidates at the level of density functional theory (DFT), both as infinite bulk monolayers and in different nanoribbon geometries. For nanoribbons, we highlight the emergence of robust metallic states whose eigenvalues cross the region of the bulk gap, and investigate the robustness of their transport properties in the framework of the non-equilibrium Green's functions (NEGF) formalism \cite{Brandbyge02}, with full account of spin-orbit interactions.
Our calculations show that naturally occurring native defects at the edge can spontaneously acquire a magnetic moment, thereby violating TRS and leading to a suppression of edge transport.
This result reveals the mechanism that is most likely to affect edge conduction in 2D topological insulators.
Interestingly, we find that chemical saturation of vacancy defects (e.g.\ by Hydrogen) is sufficient to remove the magnetic moment, thereby providing a strategy to restore topological protection at the edge.
Perhaps more surprisingly, even non-magnetic impurities may be detrimental for transport.
We indeed show that multiple non-magnetic impurities may create a channel for inter-edge scattering between opposite edge states in relatively wide ribbons, thereby affecting transport properties even when they do not represent a threat to transport individually.

The structure of this paper is as follows.
In Section \ref{sec:bandstructure} we illustrate the family of topological materials studied in this work and present their band structure in the zigzag nanoribbon geometry.
In the subsequent Sections \ref{sec:edge_defect} and \ref{sec:bulk_defect} we calculate the transport properties of topological nanoribbons in the presence of impurities, and discuss the defect-induced intra-edge and inter-edge backscattering mechanisms.
A similar formalism is then applied to zigzag MoS$_2$ nanoribbons in Section \ref{sec:non-topological}, thereby providing an insightful comparison with the transport properties of topologically trivial edge states.
Finally, Section \ref{sec:conclusions} summarizes the main results of this paper.
Appendix \ref{sec:methods} is dedicated to technical details about DFT calculations.

\section{Electronic structure of topological nanoribbons}
\label{sec:bandstructure}

The aim of this work is to address the problem of topologically protected transport by going beyond the simple model approximation, and to perform full transport calculations of realistic topological compounds from first principles. 
Thanks to the application of automated high-throughput methods in the context of material science, the portfolio of theoretically proposed 2D materials is nowadays expanding at a remarkable pace \cite{Mounet18,Haastrup18}.
This has lead to the identification of several new candidates for 2D topological insulators \cite{Marrazzo18,Olsen19,Marrazzo19_Z2}, some of which are now being tested in the lab \cite{Kandrai19,Cucchi19,Marrazzo19_dual_topo}.

We choose here to focus on Bismuth and Antimony halides, i.e. binary compounds BiX and SbX with X = (F, Cl, Br, I), whose topological nature has already been investigated in earlier work \cite{Song14_BiX_SbX,Liu14_BiX,Olsen19}.
The reason for our choice is threefold:
\begin{itemize}
	\item[(i)] They are simple binary compounds of Bi/Sb and a halogen element, with a rather small number of valence electrons per unit cell. This allows us to deal with bigger devices without extreme computational effort (compared to other candidates). 
	\item[(ii)] They are dynamically stable and thermodinamically meta-stable, in the sense that their heat of formation is negative with respect to the pure elemental form and not much larger than other competing phases (see Ref.~\cite{Haastrup18} for further discussion about thermodynamic stability of 2D materials). This means that the compound might be synthesizable, although this is not an essential aspect of our work. Indeed, we expect our conclusions to be universally applicable to any QSHI.
	\item[(iii)] Their bandgap is predicted to be in the range $0.4-1.0$ eV, which is significantly larger than most other candidates and allows for an easy distinction between topologically protected transport mediated by in-gap edge states and trivial bulk transport.
\end{itemize} 

\begin{figure*}
	\centering
	\includegraphics[width=0.8\linewidth]{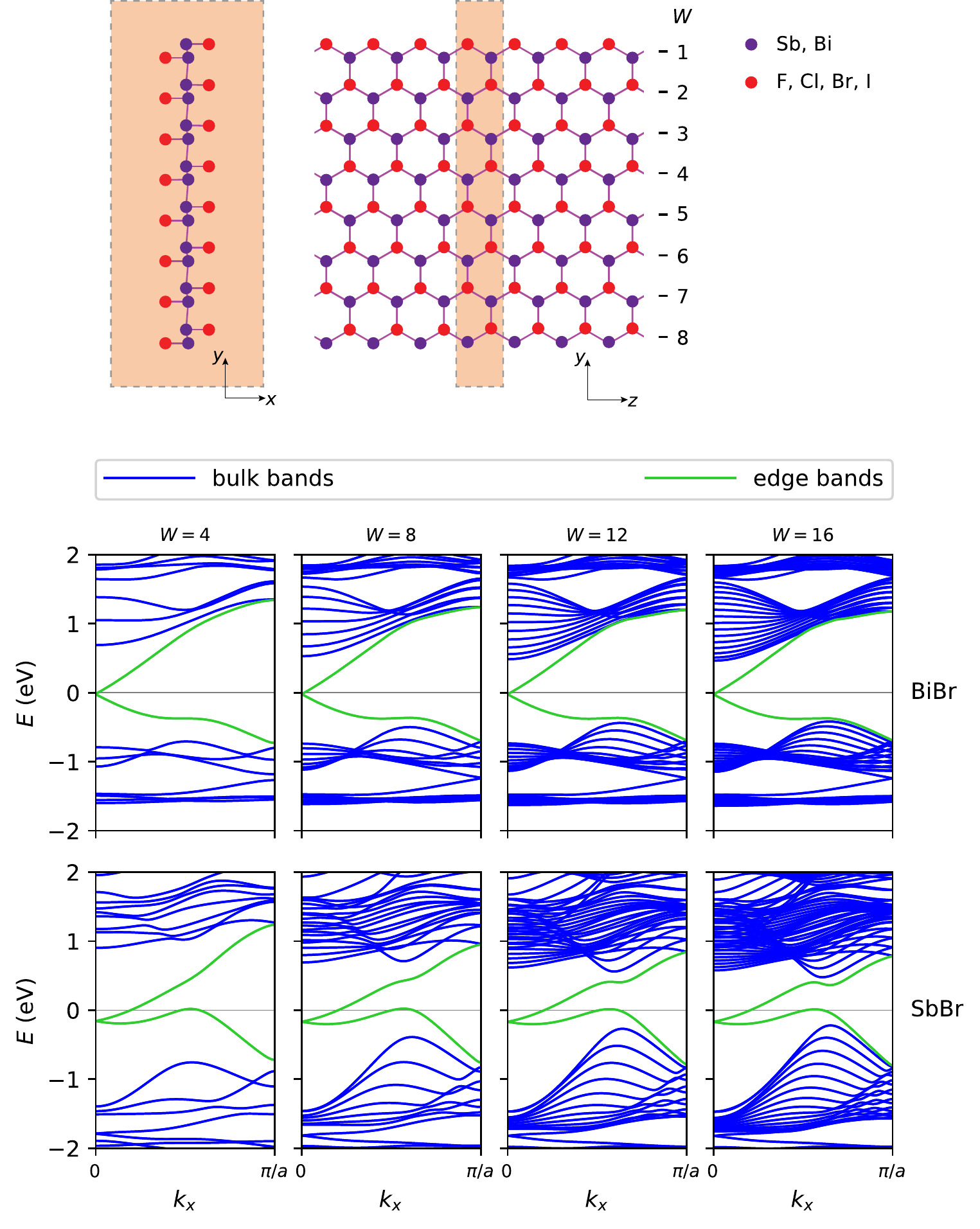}
	\caption{
		Top: a Bismuth or Antimony halide zigzag nanoribbon of width $W=8$. Dark purple dots denote the positions of Bi or Sb atoms, while halogen atoms (F, Cl, Br or I) are denoted in red. The shaded region corresponds to the unit cell used to represent the infinite ribbon.
		Bottom: Band structure of BiBr and SbBr zigzag nanoribbons of different width $W$. Topological edge states are highlighted in green. Energy is measured with respect to the Fermi energy.}
	\label{fig:zig_bands}
\end{figure*}

Monolayer Bismuth (Antimony) halides consist of a hexagonal lattice of Bi (Sb) atoms sandwiched between two layers of halogen atoms (see Fig.\ \ref{fig:zig_bands}). Halide atoms are disposed above and below the central layer in an alternate fashion, so that each Bismuth (Antimony) is bonded to one halogen only --- a similar structure occurs for the so-called \emph{graphane} \cite{Sofo07,Elias09}.
Lattice parameters and band structures for infinite 2D monolayer are all reported in the Supplemental Material \cite{Supp_Mat}.

Being band insulators with non-trivial topological invariant $\mathbb Z_2 = 1$ \cite{Song14_BiX_SbX,Olsen19}, the interface between BiX/SbX and a trivial insulator (e.g.\ vacuum) should host a pair of helical edge states. 
We confirm this by investigating the electronic structure of zigzag-terminated ribbons of different widths, such as the one shown in Fig.\ \ref{fig:zig_bands}.
Since materials with the same group-15 element but different halogens behave in a qualitatively similar fashion, we will focus on BiBr and SbBr, and show results for the remaining materials in the Supplemental Material \cite{Supp_Mat}.

Figure \ref{fig:zig_bands} shows the band structure of BiBr and SbBr zigzag nanoribbons across half of the 1D Brillouin zone.
In stark contrast with infinite 2D structures, all nanoribbons are gapless and show robust metallic states lying the region of the bulk gap, which is of the order of 0.8 eV for BiBr and 0.4 eV for SbBr.
Such states, highlighted in green in Fig.\ \ref{fig:zig_bands}, are remarkably stable as the width of the ribbon is increased, as opposed to the remaining valence and conduction states which become more and more dense and are displaced in energy as we move towards larger ribbons. We thus conclude that metallic in-gap states are localized on the edge, a fact which is also confirmed by inspecting the corresponding wavefunction (see Fig.\ S4 in Supplemental Material \cite{Supp_Mat}).
Note that the spectrum is spin degenerate owing to the presence of inversion symmetry, but metallic bands with opposite spin are located on opposite edges.

It is worth noticing that the presence of edge states is attributed to the non-trivial topological character of the bulk bands, and not to the particular edge termination. Indeed, we observe qualitatively similar features in both zigzag and armchair nanoribbons, with the latter reported in the Supplemental Material for completeness \cite{Supp_Mat}.

\section{Breakdown of edge transmission due to magnetic edge defects}
\label{sec:edge_defect}

To calculate edge transport properties of Bismuth and Antimony halides we make use of the software package QuantumATK \cite{Smidstrup17,Smidstrup19}, which allows to simulate transport devices in the framework of the NEGF formalism \cite{Brandbyge02}.
We will focus on the zero-bias transmission spectrum (TS) of a two-terminal device over an energy window of 3 eV that includes the bulk gap. This is straightforwardly linked to the two-terminal conductance of a real device through the well known Landauer formula $G = \frac{e^2}{h} T(E_\mathrm{F})$ \cite{Nazarov}.

\begin{figure*}
	\centering
	\includegraphics[width=0.8\linewidth]{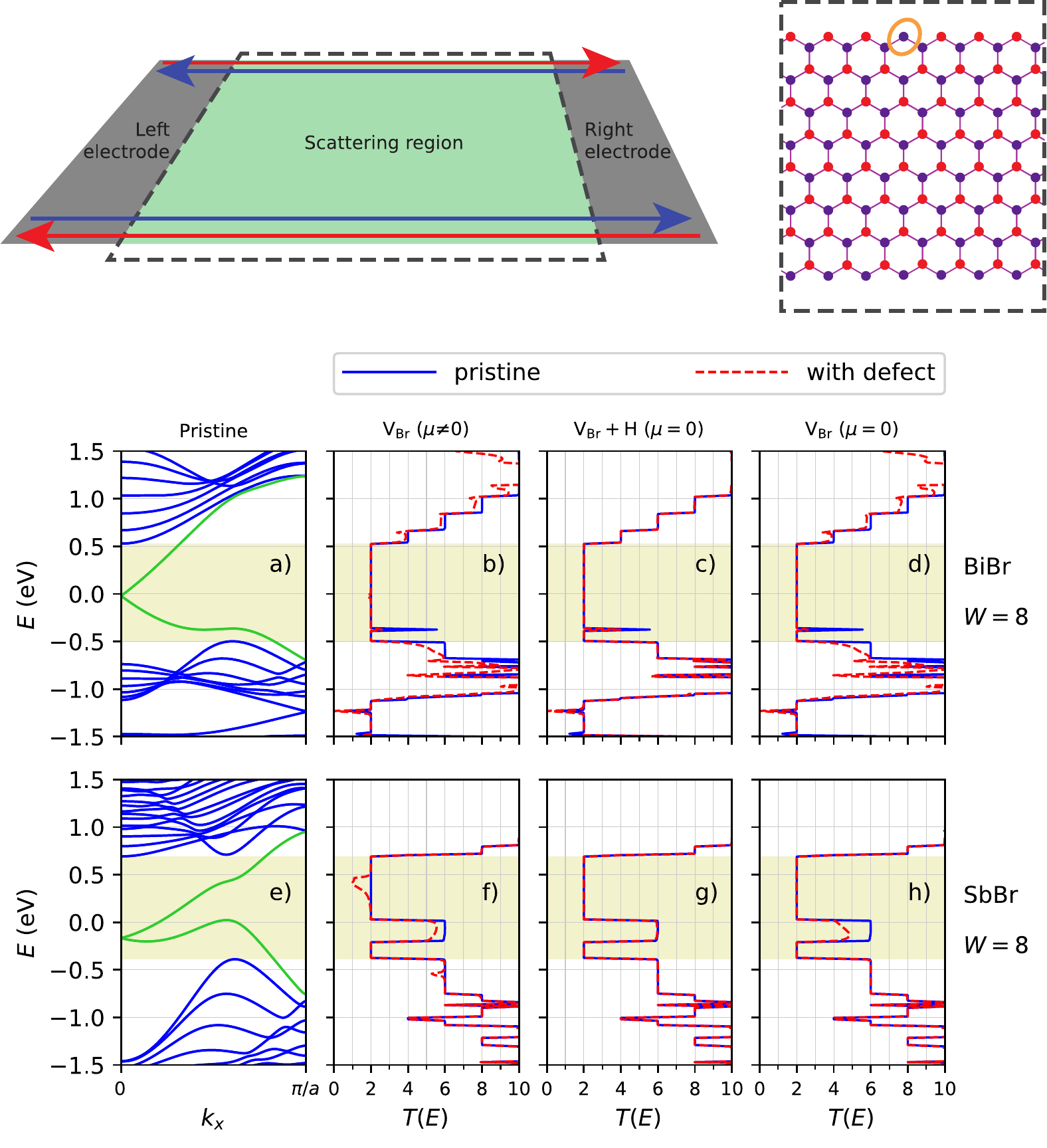}
	\caption{
	Top: Sketch of the setup used for transport calculations. Two semi-infinite electrodes (in gray) are connected to the scattering region in the middle (light green). The latter is a finite-size nanoribbon with one or more vacancy defects. A central region with an edge defect is shown to the right.
	Bottom: Transmission spectrum (TS) of BiBr and SbBr zigzag nanoribbons of width $W=8$ in presence of one of the following edge defects: magnetic edge defect (b and f); Hydrogen-saturated edge defect (c and g); non-magnetic edge defect (d and h). The TS for a pristine ribbon is also shown for comparison in each panel, and its band structure is reported in panels a and e from Fig.\ \ref{fig:zig_bands}.
	The energy region of insulating bulk is highlighted in yellow.
	}
	\label{fig:TS_edge}
\end{figure*}

Our setup is illustrated in Fig.\ \ref{fig:TS_edge} and is made as follows. 
For any given material, we create two identical, semi-infinite, pristine electrodes by repeating the nanoribbon unit cell shown in Fig.\ \ref{fig:zig_bands}.
We then create a central scattering region by considering a finite-length portion of TI nanoribbon, and remove one or more halogen atoms to account for the presence of vacancy defects (denoted as $\mathrm{V_{X}}$, X being the missing atom).
By connecting together left electrode, central scattering region and right electrode we obtain a two-terminal device setup for transport calculations.

Let us first focus on the case of a halogen vacancy on the edge (hereafter named \emph{edge defect}), as shown in Fig.\ \ref{fig:TS_edge}.
Transmission spectra for $W=8$ zigzag nanoribbons with edge defects are reported with red dashed lines in Fig.\ \ref{fig:TS_edge}, where we also report the band structure and TS of pristine ribbons.

As shown in Fig.\ \ref{fig:TS_edge}b, introducing an edge defect in BiBr does not lead to any observable effect in the region of edge states dispersion, where the transmission exactly equals the number of bands.
On the other hand, it does give a partial suppression of transmission for bulk valence and conduction states. Such a behavior is indeed the hallmark of 2D TIs, whose edge states conduction is protected even in the presence of disorder as long as time reversal symmetry is not broken.
However, the same consideration does not hold for the case of SbBr (see Fig.\ \ref{fig:TS_edge}f), which shows an unexpected and well pronounced anti-resonance in the TS around 0.4 eV and therefore a failure of topological protection. A similar feature, although much less pronounced, can be also observed around $-0.5$ eV.

To investigate the origin of the transmission dip we have checked the distribution of magnetic moments for the entire device. It shows that the configuration is almost entirely non-magnetic, except for a magnetic moment of $0.9 \mu_\mathrm{B}$ which is exactly localized at the Sb site underneath the Br vacancy. On the other hand, the BiBr nanoribbon have a negligible small magnetic moment at the Br vacancy.
We thus attribute the suppression of edge conductance in SbBr nanoribbons to the spontaneous magnetization of the edge impurity, which invalidates topological protection and allows for intra-edge back-scattering.
This is a recurrent feature for all Antimony halides, for which we systematically observe an anti-resonance in a narrow energy window around 0.2--0.5 eV due to the formation of localized magnetic moments (see Supplemental Material \cite{Supp_Mat}).

It is interesting to notice that a similar mechanism has been proposed very recently in the framework of the Kane-Mele-Hubbard model in graphene, where the breakdown of time reversal symmetry at vacancy defects is shown to lead to a corresponding breakdown of conductance quantization \cite{Novelli19}. The present \emph{ab-initio} calculations support this picture and also bear similarities with simple theoretical models in which magnetic scatterers, such as magnetic adatoms or ferromagnetic gates, are introduced in 2D TIs \cite{Dang16,Zheng18}.

The magnetization of edge defects originates from the presence of a dangling bond at the vacancy site, which makes the configuration chemically unstable and drives the formation of a localized magnetic moment, as reported in earlier work \cite{Nair13,Khan17,He10,Li16,Chen11,Azevedo09}. We thus conjecture that the chemical saturation of the dangling bond with a suitable element should eliminate any magnetic structure at the defect and restore a perfect transport at the edge. Indeed, we find that saturation with Hydrogen restores the perfect transmission, as shown in Figs.\ \ref{fig:TS_edge}c and \ref{fig:TS_edge}g.

We have also calculated the TS for a non-magnetic configuration of the halogen vacancy (we manually set all magnetic moments to zero), which we report in Figs.\ \ref{fig:TS_edge}d and \ref{fig:TS_edge}h. As expected, there's no backscattering in this case since time reversal symmetry is not violated. However, we find rather small energy difference between the magnetic and non-magnetic configurations, which might be beyond the accuracy of DFT calculations.

\begin{figure*}
	\centering
	\includegraphics[width=0.8\linewidth]{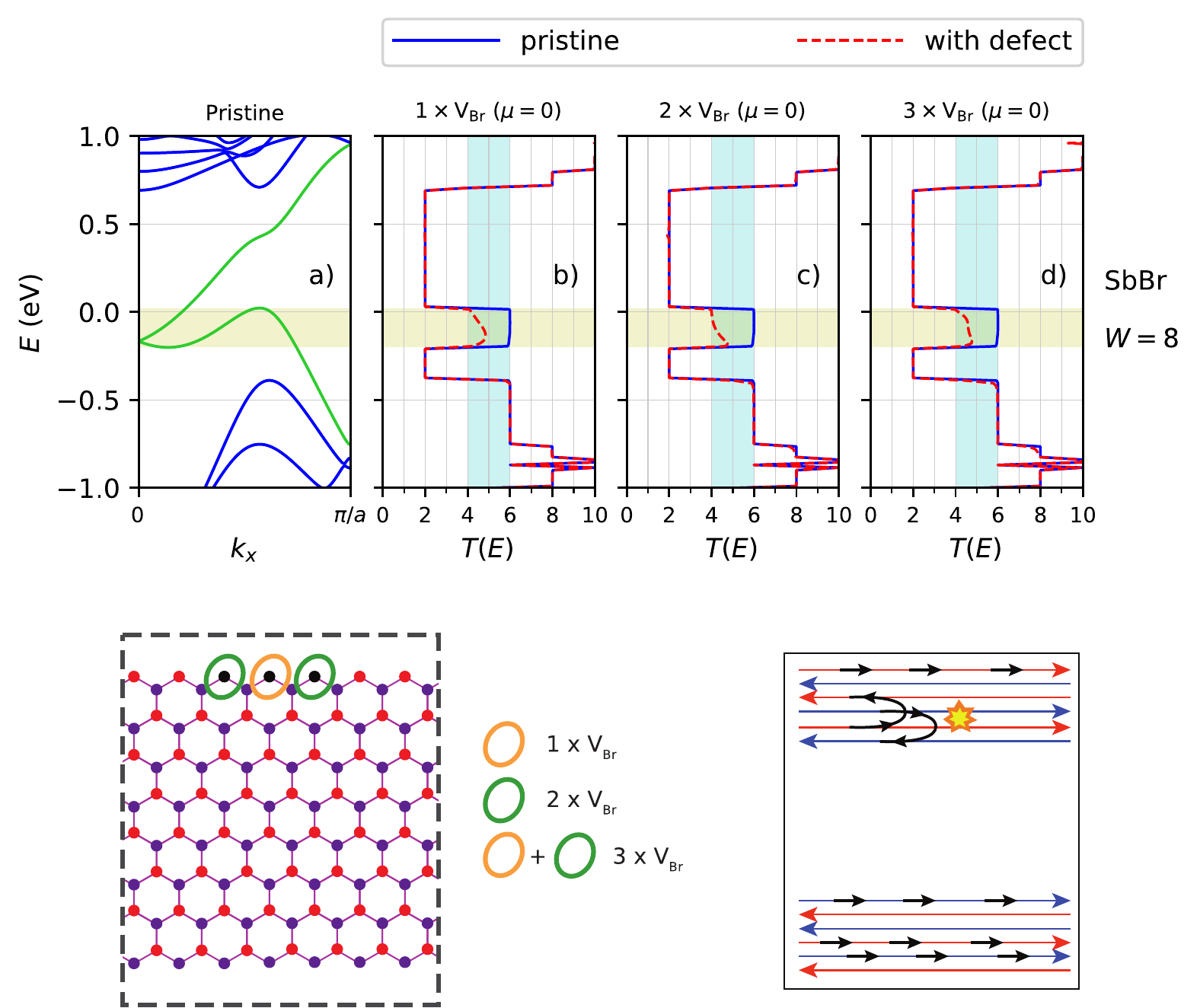}
	\caption{
		Transmission spectrum (TS) of an SbBr zigzag nanoribbon of width $W=8$ in presence of multiple non-magnetic edge defects as shown in the bottom panel: single edge defect, reported from Fig.\ \ref{fig:TS_edge}h (b); double edge defect (c); triple edge defect (d).
		The TS for a pristine ribbon is also shown for comparison in each panel, and its band structure is reported from Fig.\ \ref{fig:zig_bands} in panel a.	
		The energy region in which the nanoribbon bears 3 pairs of edge states in highlighted in yellow, while the region $4 \leq T(E) \leq 6$ is highlighted in cyan.
		A graphical interpretation of the results in terms of open and closed channels on each edge is given in the bottom right panel, where metallic edge states with opposite spin polarization are represented with different colors (red and blue).
	}
	\label{fig:TS_edge_multiple}
\end{figure*}

There is however one last puzzling question emerging from Fig.\ \ref{fig:TS_edge}, which is the unexpected back-scattering observed in SbBr nanoribbons between $-0.2$ and 0.0 eV, that is where $T(E) = 6$ (see panels f and h). We attribute this result to the following mechanism. Due to the presence of non-monotone energy dispersion of the edge states, the nanoribbon actually hosts three pairs of metallic states per edge in this energy range, whose direction of motion can be easily inferred from the slope of the bands.\footnote{A similar effect, although much less pronounced, occurs for BiBr around $-0.4 \mathrm{eV}$.} Thus, it becomes possible for an electron to scatter into a state with same spin but different direction of motion, as schematically depicted in Fig.\ \ref{fig:TS_edge_multiple}. 
This mechanism can never lead to the total suppression of edge conductance, as there will always be a pair of helical edge states which are not accompanied by the corresponding counter-propagating states. In other words, in the presence of impurities or disorder, whatever odd number of edge states pairs is practically equivalent to a single pair, which is a manifestation of the binary nature of the $\mathbb Z_2$ topological invariant.
This implies that the TS in the region between $-0.2$ and 0.0 eV may approach $T(E)=4$ as the impurities become more and more abundant, since the three transport channels on the bottom edge will be accompanied by only one surviving pair on the disordered top edge.
To check this, we have investigated different non-magnetic configurations of SbBr nanoribbons with multiple edge defects, which are all reported in Fig.\ \ref{fig:TS_edge_multiple} together with the single-impurity configuration previously discussed.
Indeed, the resulting TS never drops below 4 in the aforementioned region, supporting our interpretation.

In passing, it is worth noting that we have also calculated the TS for smaller ribbons ($W=4$) and for all remaining materials in the presence of similar edge defects, obtaining qualitatively similar results --- that is, sharp anti-resonance in the conductance due to the formation of magnetic defects, and intra-edge backscattering when the structure carries three helical pairs per boundary. These results are shown in the Supplemental Material \cite{Supp_Mat}.

\section{Inter-edge scattering mediated by non-magnetic bulk defects}
\label{sec:bulk_defect}

We now turn our attention to halogen vacancies located away from the edge, which will be denoted \emph{bulk defects}.
As for the case of edge defects, we have studied both magnetic and non-magnetic configurations. However, we will only present results for non-magnetic bulk defects in BiBr for the sake of clarity.

\begin{figure*}
	\centering
	\includegraphics[width=0.8\linewidth]{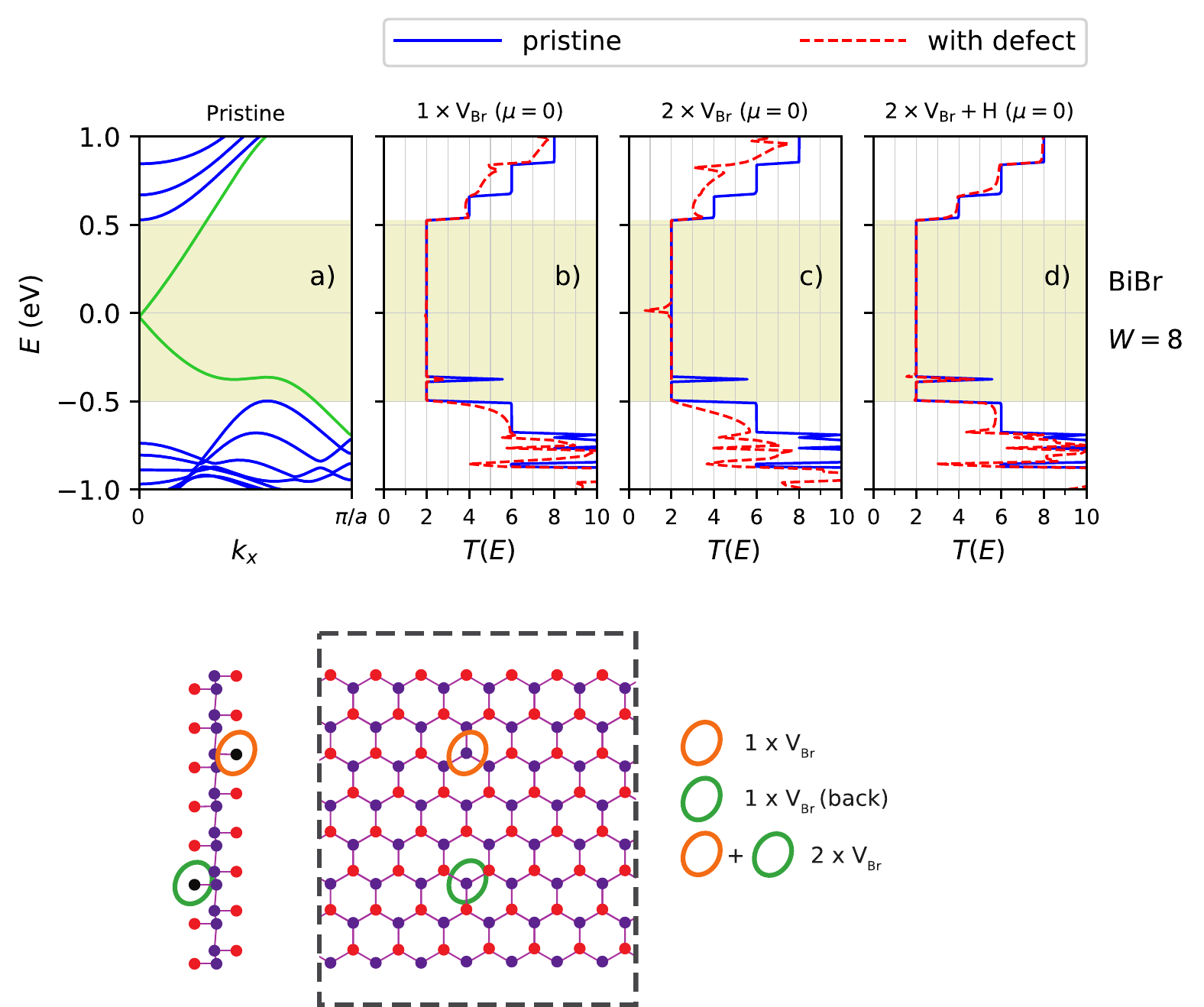}
	\caption{
		Transmission spectrum (TS) of a BiBr zigzag nanoribbon of width $W=8$ in presence of multiple non-magnetic bulk defects as sketched in the bottom panel: single bulk defect (b); double bulk defect (c); Hydrogen-saturated double bulk defect (d).
		The TS for a pristine ribbon is also shown for comparison in each panel, and its band structure is reported in panel a from Fig.\ \ref{fig:zig_bands}.
		The energy region of insulating bulk is highlighted in yellow.
	}
	\label{fig:TS_bulk_multiple}
\end{figure*}

We have explored a scenario where two halogen atoms are simultaneously removed from a BiBr nanoribbon of width $W=8$, leaving a couple of bulk vacancy defects in the central region.
As shown in Fig.\ \ref{fig:TS_bulk_multiple}, both vacancies are symmetrically placed at a distance $d=\sqrt{3}a$ from the edge, with the defect-defect distance being $d'=5a/\sqrt{3}$ (i.e.\ $d=9.46$ \AA\ and $d'=15.76$ \AA\ for BiBr).

The TS of such a configuration is reported in Fig.\ \ref{fig:TS_bulk_multiple}. While both impurities do not affect transport properties in the bulk gap region individually, which is demonstrated by the perfect TS in the region $-0.5 \mathrm{eV} \lesssim E \lesssim 0.5 \mathrm{eV}$ in Fig.\ \ref{fig:TS_bulk_multiple}b, they do suppress transport at energy $E \approx 0$ when they are simultaneously present, which is reflected in the sharp anti-resonance in Fig.\ \ref{fig:TS_bulk_multiple}c.

\begin{figure*}
	\centering
	\includegraphics[width=0.8\linewidth]{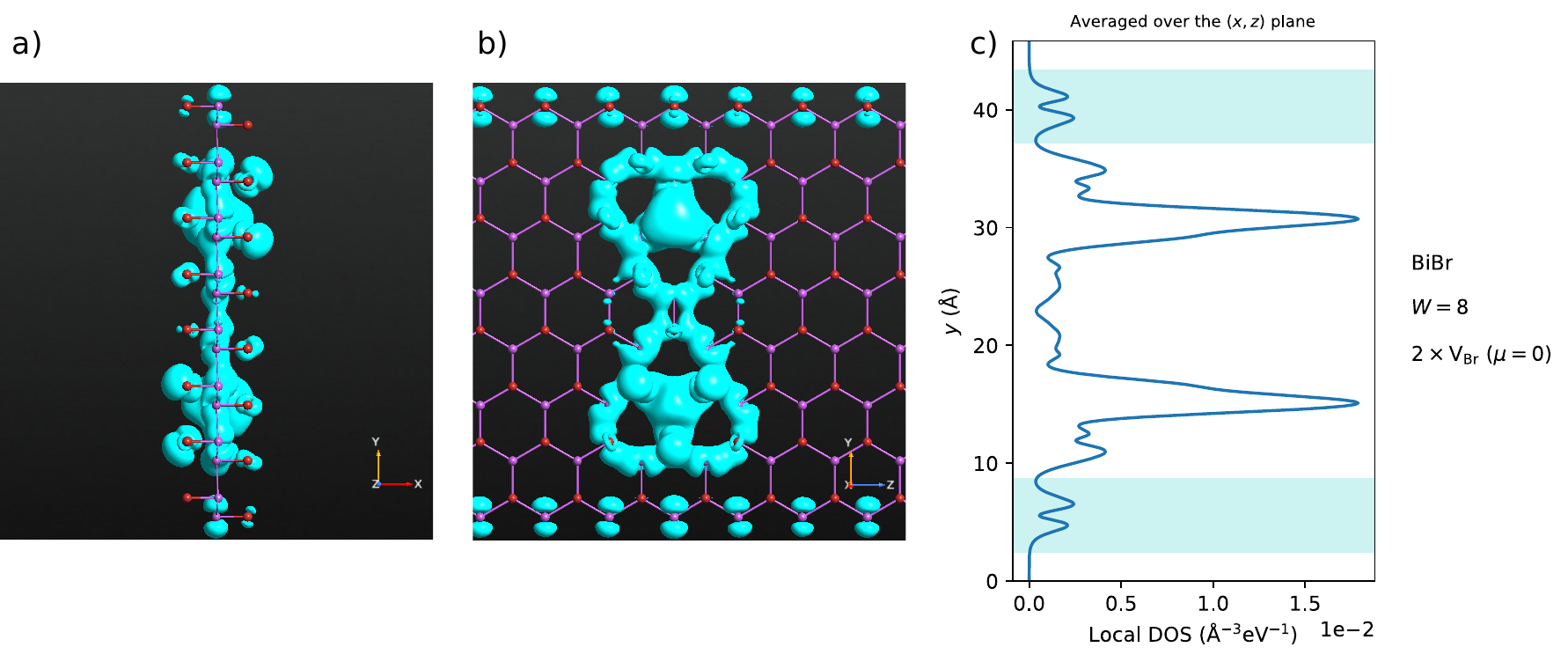}
	\caption{Local density of states (LDOS) for the configuration in Fig.\ \ref{fig:TS_bulk_multiple}. Side and top views of the surface $\mathrm{LDOS}(x,y,z) = 0.02 \,\mathrm{\AA}^{-3} \mathrm{eV}^{-1}$ are shown in panels a and b respectively. Panel c shows the LDOS in the transverse direction $y$ averaged over the $xz$ plane, with the spatial region spanned by the edge states highlighted in cyan.}
	\label{fig:LDOS}
\end{figure*}

The corresponding scattering event is of inter-edge nature, as intra-edge scattering is forbidden by TRS (we are considering a non-magnetic structure). An electron traveling along the top boundary from left to right can hop on the closest defect state, whose energy actually lies in the bulk gap range. From there, it has a finite possibility of reaching the second impurity, due to a non-zero matrix element between localized states at the impurities. Finally, it tunnels into the bottom edge states, where it propagates back towards the left electrode without having to flip the spin polarization.
This analysis is further confirmed by the behavior of the local density of states shown in Fig.\ \ref{fig:LDOS}, which clearly demonstrates that the impurity states occupy a large transverse portion of the nanoribbon, while also having a substantial overlap between them.
In this scenario, it is the co-operation between multiple impurities that creates a path for inter-edge scattering, even in such cases where they would not represent any threat to transport if considered individually.
Such a mechanism has been frequently neglected in the literature, which rather focus on the effect of single impurities on the transport properties of 2D TIs. Nevertheless, it could play a major role when the defect concentration exceeds a certain threshold \cite{Tiwari19,Chang14}, or when opposite edge states are deliberately funneled through a narrow constriction \cite{Strunz20}.

It is worth noting that chemical saturation of the dangling bonds with Hydrogen removes the energy levels of the defects from the bulk gap region. The corresponding TS, which we show in Fig.\ \ref{fig:TS_bulk_multiple}d, is basically unaffected, thus suggesting a strategy to minimize the impact of impurity-mediated inter-edge scattering on transport.

Finally, we mention that we have observed signatures of inter-edge scattering for different disordered configurations and nanoribbon widths, as reported in the Supplemental Material \cite{Supp_Mat}.
In particular, opposite edge states in narrow nanoribbons ($W=4$) can be coupled together by one single bulk impurity, since the spatial extension of the defect wavefunction becomes comparable with the nanoribbon width \cite{Supp_Mat}.

\section{Comparison with non-topological edge states}
\label{sec:non-topological}

So far, we have discussed two mechanisms that lead to the failure of conductance quantization in 2D TIs.
In this section, we illustrate how the results presented up to now are clearly related to topology, and how edge states would behave in the absence of topological protection.

\begin{figure*}
	\centering
	\includegraphics[width=0.8\linewidth]{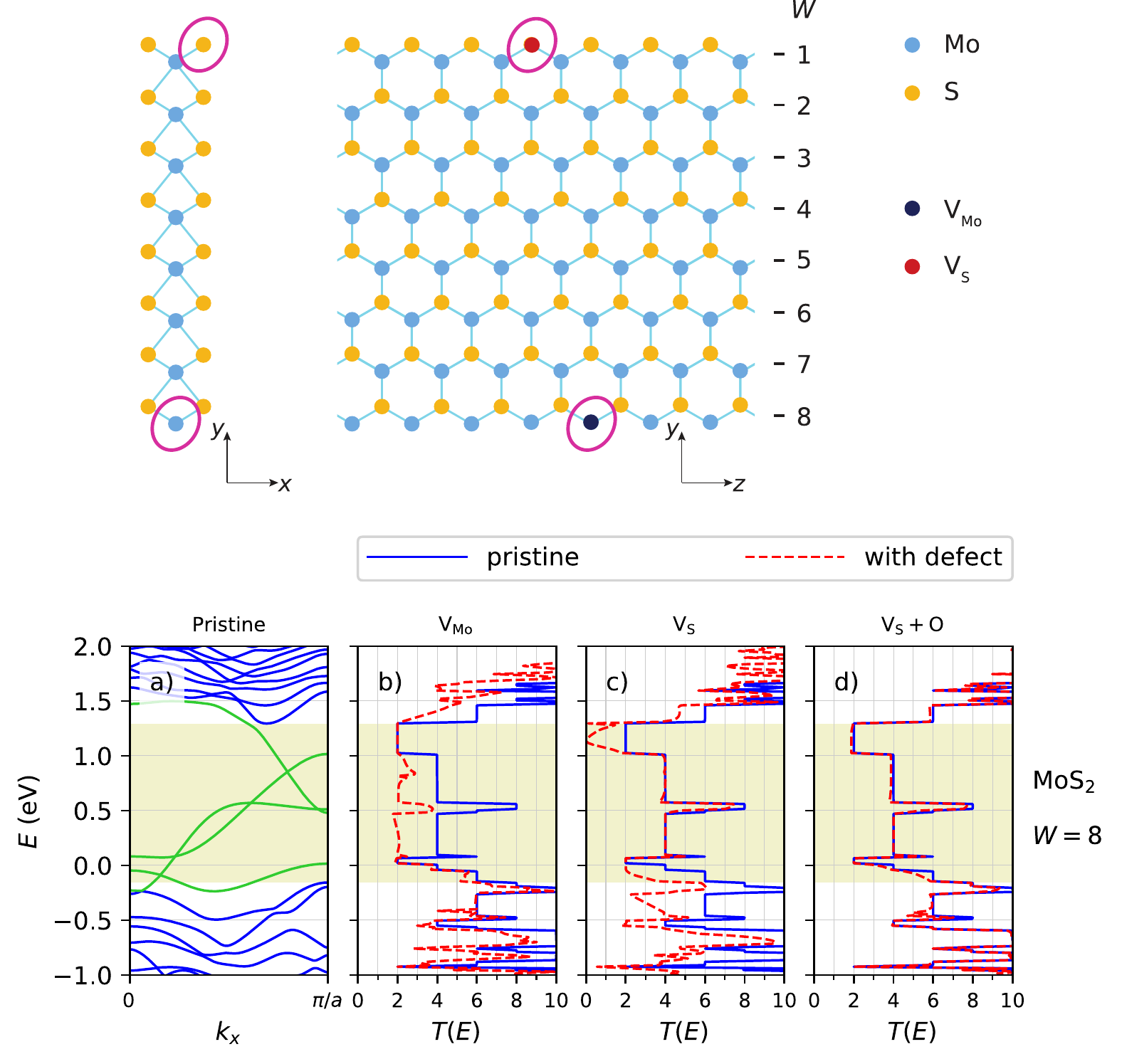}
	\caption{
		Top: Top and lateral view of an MoS$_2$ zigzag nanoribbon of width $W=8$. Atoms removed to create defects are denoted with different colors.
		Bottom: Transmission spectrum (TS) of MoS2 zigzag nanoribbons of width $W=8$ in presence of one of the following edge defects shown in Fig.\ \ref{fig:TS_edge_MoS2}: Mo vacancy (b); S vacancy (c); Oxygen-saturated S vacancy (d). The TS for a pristine ribbon is also shown for comparison in each panel, and its band structure is shown in panel a. The energy region of insulating bulk is highlighted in yellow.}
	\label{fig:TS_edge_MoS2}
\end{figure*}

We thus focus here on the transport properties of a MoS$_2$ nanoribbon, whose crystal structure is shown in Fig.\ \ref{fig:TS_edge_MoS2}.
Such structures are known to host metallic edge states at the zigzag termination which are however not due to topology, but rather to polar discontinuity at the interface between MoS$_2$ and the vacuum \cite{Gibertini14,Gibertini15}.
We will therefore call these \emph{trivial edge states}, to indicate that they are not generated as boundary states between materials with different topological invariants.

In Fig.\ \ref{fig:TS_edge_MoS2}a we show the band structure of a pristine MoS$_2$ nanoribbon of width $W=8$. The metallic edge states, indicated with a different color in the figure, are clearly visible.
We then create two different defect configurations by removing an atom on the edge (either Mo or S).
We calculate the TS for such structures neglecting the contribution from SOC, which is known to have a negligible effect in this case, and compare them to the TS of a pristine nanoribbon.

Figures \ref{fig:TS_edge_MoS2}b and \ref{fig:TS_edge_MoS2}c show that the presence of an edge defect has a dramatic consequence for the transmission properties of the nanoribbon.
Generally, we observe a much larger effect of backscattering as compared to the case of topologically protected materials.
The conductance is basically halved over the entire energy range $0.0-1.0$eV when we remove a single Mo atom from the edge, and is completely blocked in the range $1.0-1.3$eV when the edge defect is a Sulfur vacancy.
We conclude that the trivial edge states of MoS$_2$ nanoribbons are much more prone to backscattering than the topologically protected edge states in the Bi (Sb) halides.

Finally, we have also calculated the TS for the case of Oxygen-saturated S vacancy, which is shown in Fig.\ \ref{fig:TS_edge_MoS2}d.
Once again, chemical saturation seems to be beneficial to edge transport, as we recover an almost perfect transmission. However, the TS still deviates by 3\% from the pristine value in the region $0.5-1.3$ eV. In contrast, the TS for H-saturated edge defects in BiBr and SbBr nanoribbons never deviates more that 0.1\% from the corresponding quantized values.

\section{Conclusions}
\label{sec:conclusions}

\begin{figure}
	\centering
	\includegraphics[width=\linewidth]{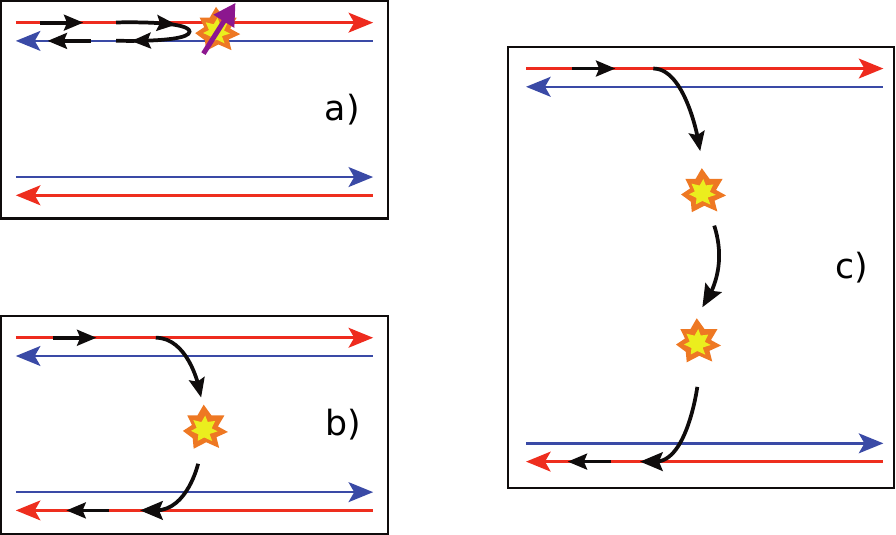}
	\caption{Graphical representation of different backscattering mechanisms: intra-edge scattering due to magnetic edge impurities (a); inter-edge scattering through a single non-magnetic bulk defect (b); inter-edge scattering mediated by multiple bulk impurities (c). Metallic edge states with opposite spin polarization are represented with different colors (red and blue).}
	\label{fig:discussion}
\end{figure}

In summary, we have performed \emph{ab-initio} transport calculations of two-terminal TI nanoribbons using the NEGF formalism. By accounting for the presence of both edge and bulk defects, we have pinpointed two sources of backscattering which lead to the breakdown of conductance quantization:
\begin{itemize}
	\item[(i)] \emph{Intra-edge scattering due magnetic edge impurities.} The dangling bond originated by the removal of one atom at the edge may in some cases drive the formation of a localized magnetic moment at the impurity site. This local breaking of TRS allows for backscattering events involving spin-flip (as sketched in Fig.\ \ref{fig:discussion}a).
	This basically blocks electrical conduction through one of the edges, while leaving the opposite one unperturbed. The corresponding transmission function drops from 2 to 1 when the energy of the incoming electron resonates with the defect level.
	\item[(ii)] \emph{Inter-edge scattering due to (multiple) bulk impurities.} In a narrow nanoribbon, a single non-magnetic bulk impurity can open a backscattering channel between opposite edge states without breaking TRS (Fig.\ \ref{fig:discussion}b). Although this can be obviously avoided by increasing the nanoribbon width, large nanoribbons will be still affected by inter-edge scattering above a certain threshold of defect concentration, when multiple bulk impurities generate a backscattering path across the structure (Fig.\ \ref{fig:discussion}c).
\end{itemize}

Finally, it's worth mentioning that our results are by no means limited to the particular class of materials chosen in this work. Rather, we expect them to be relevant for all QSHIs.
We also anticipate that similar mechanisms may deteriorate surface conduction of three-dimensional time-reversal invariant TIs.

\begin{acknowledgments}

We thank Nicola Marzari for stimulating discussions, Mads Brandbyge and Tue Gunst for useful discussions and technical help.
The research leading to these results has received funding from the European Union's Horizon 2020 research and innovation program under the Marie Sk\l{}odowska-Curie grant agreement No.\ 754462. (EuroTechPostdoc).
KST acknowledges funding from the European Research Council (ERC) under the European Union's Horizon 2020 research and innovation programme (Grant No.\ 773122, LIMA).
The Center for Nanostructured Graphene is sponsored by the Danish National Research Foundation, Project DNRF103.

\end{acknowledgments}

\appendix

\section{Methods}
\label{sec:methods}

All structures are obtained directly from the C2DB database \cite{Haastrup18}, where they are relaxed with the Perdew-Burke-Ernzerhof (PBE) \cite{Perdew96} exchange-correlation functional using the DFT code GPAW \cite{Mortensen05,Enkovaara10} and the software package ASE \cite{HjorthLarsen17}.
For further details about this first step we refer to Ref.\ \cite{Haastrup18}.

To create a nanoribbon, we define a rectangular unit cell as shown in Fig.\ \ref{fig:zig_bands}, including a different number of atoms according to the width $W$. Each ribbon is separated from its periodic replicas by including a suitable amount of vacuum in the unit cell, both in the out-of-plane and the non-periodic in-plane directions.

We then use the atomistic simulation toolkit QuantumATK \cite{Smidstrup17,Smidstrup19} to calculate the DFT band structure of pristine nanoribbons. For QuantumATK band structure calculations we resorted to an LCAO basis using the SG15 pseudo-potentials \cite{Schlipf15} --- with the only exception of BiCl and SbCl structures, where we use the OpenMX package \cite{Ozaki03,Ozaki04}.
We sample the Brillouin zone of the nanoribbon with a $1\times1\times16$ Monkhorst–Pack (MP) grid \cite{Monkhorst76} to ensure convergence, while the density mesh cutoff controlling the real-space grid is set to 100 Hartree. 
We also checked the band structure with GPAW by using a plane-wave basis set with an energy cutoff of 400 eV and an identical MP grid, obtaining an excellent agreement with QuantumATK calculations. In both cases we use the PBE functional with the inclusion of spin-orbit coupling, which is a crucial ingredient here.

To calculate the transmission spectrum at zero bias we use the NEGF formalism \cite{Brandbyge02} as implemented in QuantumATK.
We define a transport setup by creating identical left and right electrodes and a central scattering region, as shown in Fig.\ \ref{fig:TS_edge}. We create pristine electrodes by using the same material as in the central region, and make sure that electrodes are well screened by repeating the unit cell in the transport direction ($z$ axis) a suitable amount of times --- only once for BiCl and SbCl, three times for MoS$_2$ and twice for the remaining materials.
One or more defects in the central scattering region are created by removing one or more halogen atoms. Note that structures obtained in such a way are not optimized, so that we neglect reconstruction effects.
A special care is taken in including a suitable portion of pristine material at both sides of the central scattering region, so that the latter is smoothly connected to the electrodes.

Electrodes for NEGF calculations are sampled with a $1\times1\times150$ MP grid, and we use a slightly lower mesh cutoff of 80 Hartree. 
Periodic boundary conditions are imposed to the device along the transverse directions $x$ and $y$, while we use Dirichlet boundary conditions for the transport direction $z$.

Finally, magnetic moments are estimated from a Mulliken population analysis \cite{Mulliken55}, where we take the difference between the spin up and down populations at each site as the magnitude of the magnetic moment in units of the Bohr magneton.

\bibliography{biblio.bib}

\end{document}


\title{Supplemental Material to ``Conductance of quantum spin Hall edge states from first principles: the critical role of magnetic impurities and inter-edge scattering''}
	
\author{Luca Vannucci}
\email{lucav@fysik.dtu.dk}
\affiliation{CAMD, Department of Physics, Technical University of Denmark, 2800 Kgs. Lyngby, Denmark}
	
\author{Thomas Olsen}
\affiliation{CAMD, Department of Physics, Technical University of Denmark, 2800 Kgs. Lyngby, Denmark}
	
\author{Kristian S. Thygesen}
\affiliation{CAMD, Department of Physics, Technical University of Denmark, 2800 Kgs. Lyngby, Denmark}
\affiliation{Center for Nanostructured Graphene (CNG), Technical University of Denmark, 2800 Kgs. Lyngby, Denmark}
	
\date{\today}

\maketitle

\tableofcontents

\clearpage

\section{Band structure of Antimony and Bismuth halides}
\label{subsec:bulk}

\subsection{2D monolayers}

\begin{figure}
	\centering
	\includegraphics[width=0.9\linewidth]{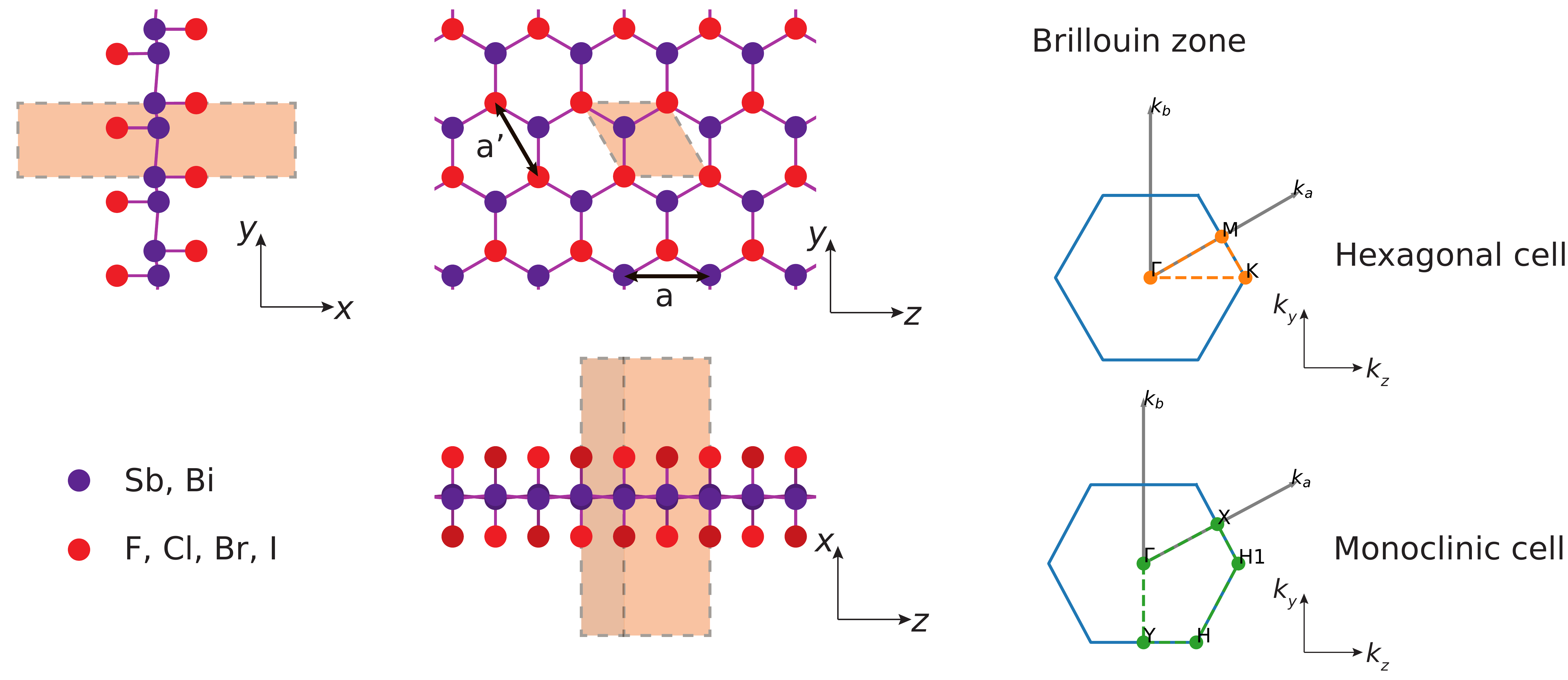}
	\caption{Lattice structure of monolayer BiX and SbX (X = F, Cl, Br, I). The unit cell is marked with a shaded region. The corresponding Brillouin zone in shown on the right, and the path used for band structure calculations is reported with dashed colored lines. We use the conventional $\mathrm{\Gamma M K \Gamma}$ path for materials with hexagonal unit cell, while the path $\mathrm{\Gamma Y H H_1 X \Gamma}$ is used for the case of monoclinic cell (such as SbI).}
	\label{fig:lattice}
\end{figure}

The crystal structure of monolayer Bismuth (Antimony) halides consists of a low-buckled layer of Bismuth (Antimony) sandwiched between two layers of halogen atoms, which are disposed in an alternated fashion with respect to the central layer (see Figure \ref{fig:lattice}). Seven structure out of 8 have hexagonal unit cell, with both inversion symmetry and a three-fold rotational symmetry. SbI is the only exception, having monoclinic structure with broken rotational symmetry.

\begin{table}
	\centering
	\begin{tabular}{lcccccc}
		material & a (\AA) & a' (\AA) & d (\AA) & h (\AA) & $\Delta_\mathrm{GPAW}$ (eV) & $\Delta_\mathrm{ATK}$ (eV) \\
		\hline
		BiF &	5.32 &		 &	2.09 &	0.42 &	0.99 &	1.02\\
		BiCl & 	5.45 &		 &	2.52 &	0.21 & 	0.90 &	0.90\\
		BiBr &	5.46 &		 &	2.67 &	0.17 & 	0.86 &	0.83\\
		BiI &	5.48 &		 &	2.87 &	0.10 & 	0.78 &	0.86\\
		SbF & 	5.10 &		 &	1.97 &	0.28 & 	0.36 &	0.34\\
		SbCl & 	5.22 &		 &	2.40 &	0.11 & 	0.43 &	0.40\\
		SbBr &	5.24 &		 &	2.57 &	0.09 & 	0.44 &	0.41\\
		SbI &	5.05 &	5.36 &	2.78 &	0.02 &	0.49 &	0.43\\
		\hline
	\end{tabular}	
	\caption{Lattice parameters and band gap of monolayer BiX and SbX (X = F, Cl, Br, I). a: hexagonal lattice constant; a': 2nd lattice constant (for materials with monoclinic structure); d: Bi-X or Sb-X distance; h: buckling height; $\Delta_\mathrm{GPAW}$: electronic band gap obtained with GPAW \cite{Enkovaara10}; $\Delta_\mathrm{ATK}$: electronic band gap obtained with QuantumATK \cite{Smidstrup17,Smidstrup19}.}
	\label{tab:params}
\end{table}

Structural parameters for all materials --- such as lattice constants $a$ and $a'$, Bi(Sb)-halogen distance $d$, buckling height $h$ --- are reported in Table \ref{tab:params}, together with the corresponding 2D band gap. Results are in good agreement with previous work \cite{Song14_BiX_SbX}.

\begin{figure}
	\centering
	\includegraphics[width=0.49\linewidth]{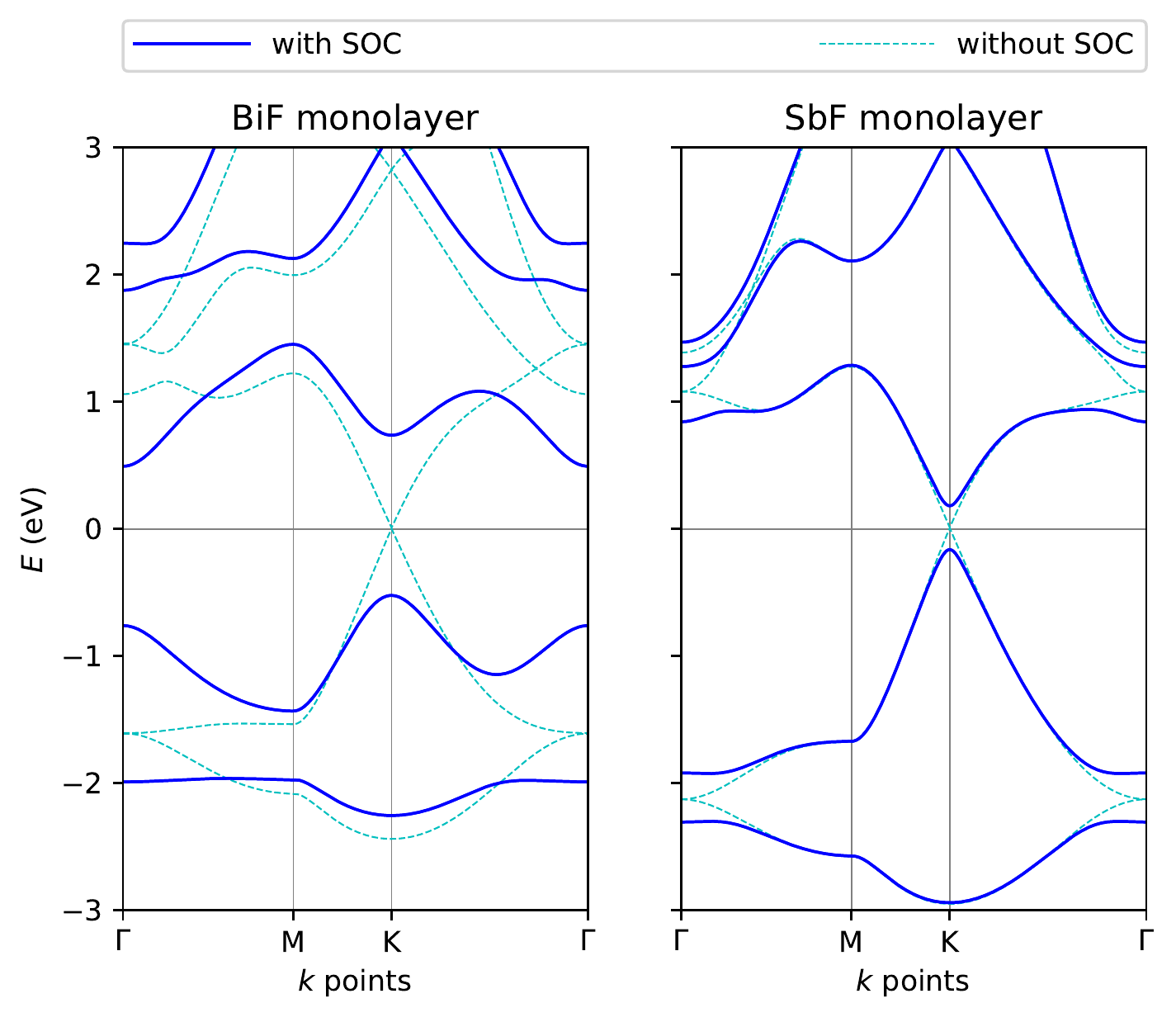}
	\includegraphics[width=0.49\linewidth]{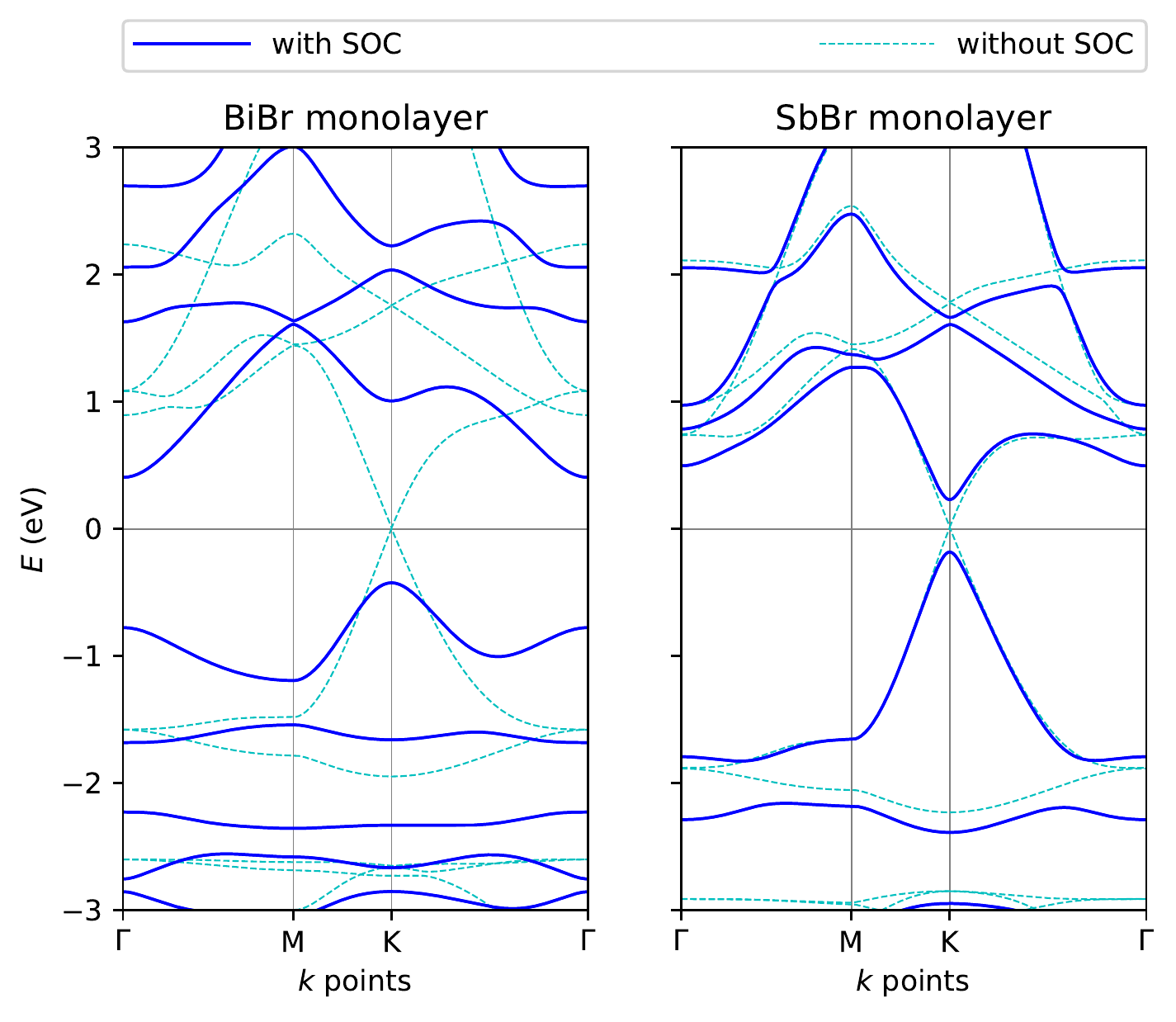}
	\includegraphics[width=0.49\linewidth]{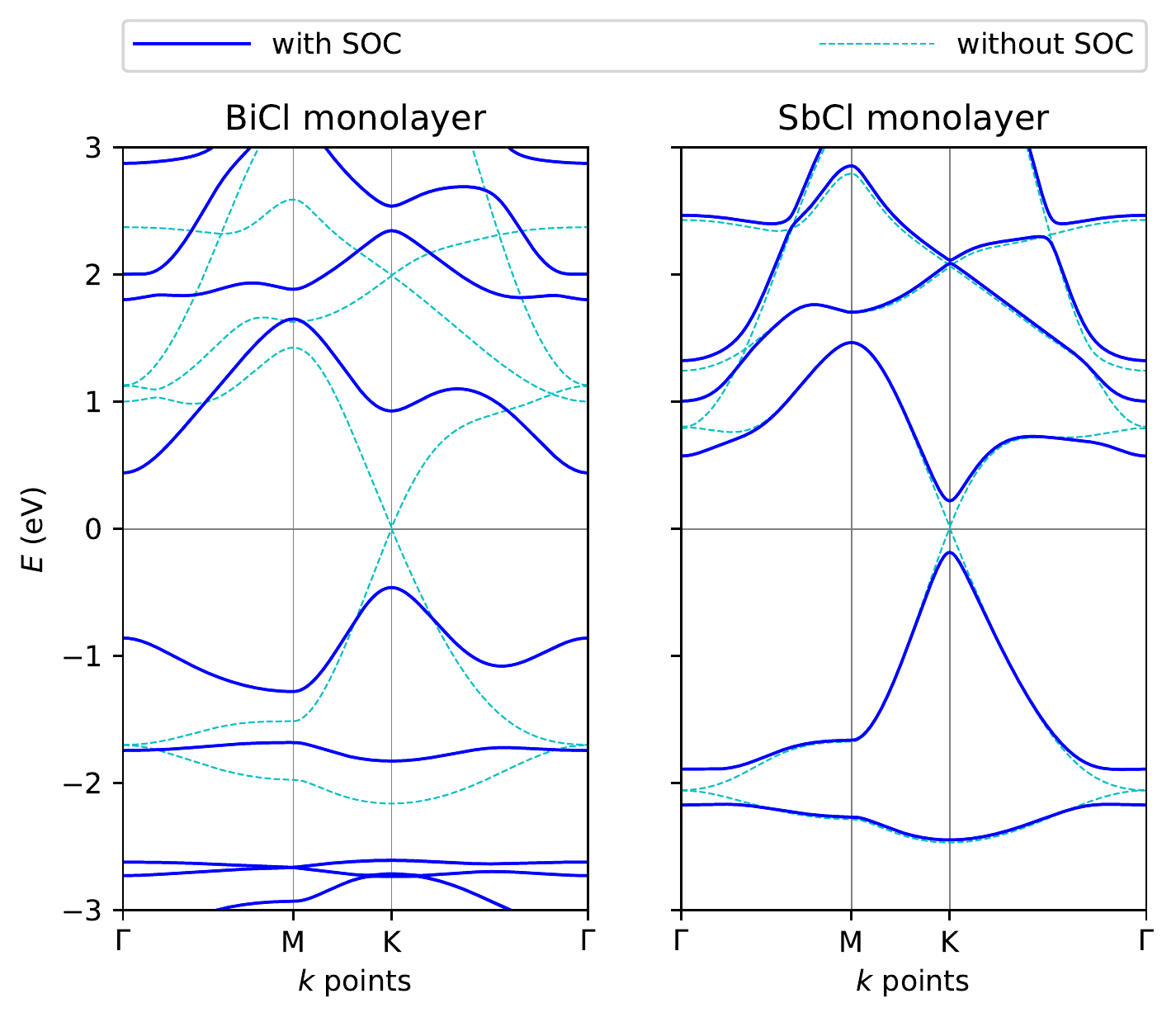}
	\includegraphics[width=0.49\linewidth]{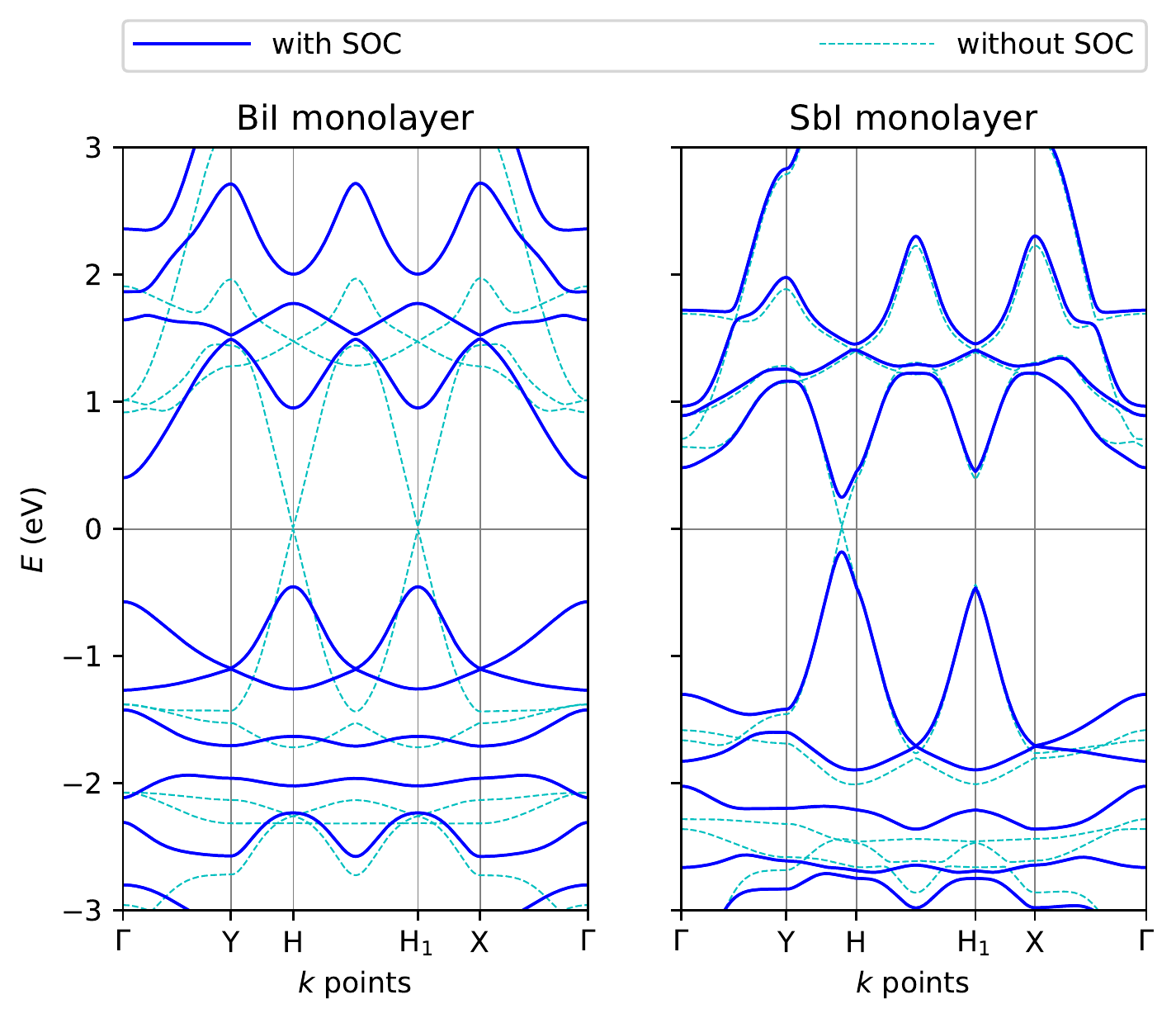}
	\caption{Band structure of bulk 2D monolayers along the path shown in Fig.\ \ref{fig:lattice}. We plot the eigenvalues both without and with the inclusion of spin-orbit coupling (SOC). Note that BiI has actually a hexagonal unit cell, but we use the monoclinic path for a better comparison with SbI.}
	\label{fig:bulk}
\end{figure}

The band structure of 2D monolayers is reported in Fig.\ \ref{fig:bulk}. Here we have calculated electronic bands both without and with spin-orbit coupling (SOC). In the absence of SOC, all materials are semi-metals with a Dirac cone at the K point (slightly shifted for the monoclinic structure SbI). However, SOC has an essential role here as it opens a gap of $0.3-0.4$eV for Antimony-based materials, and $0.8-1.0$eV for Bismuth-based materials.
The presence of a SOC-induced band gap is indeed a characteristic feature of topological materials \cite{Qi11}.

\FloatBarrier
\subsection{Zigzag nanoribbons}

Here we report the band structure of zigzag nanoribbons for all materials that have not been shown in the main text. All structures are gapless with a pair of metallic in-gap states that is quite stable with respect to the width $W$.

\begin{figure}
	\centering
	\includegraphics[width=0.49\linewidth]{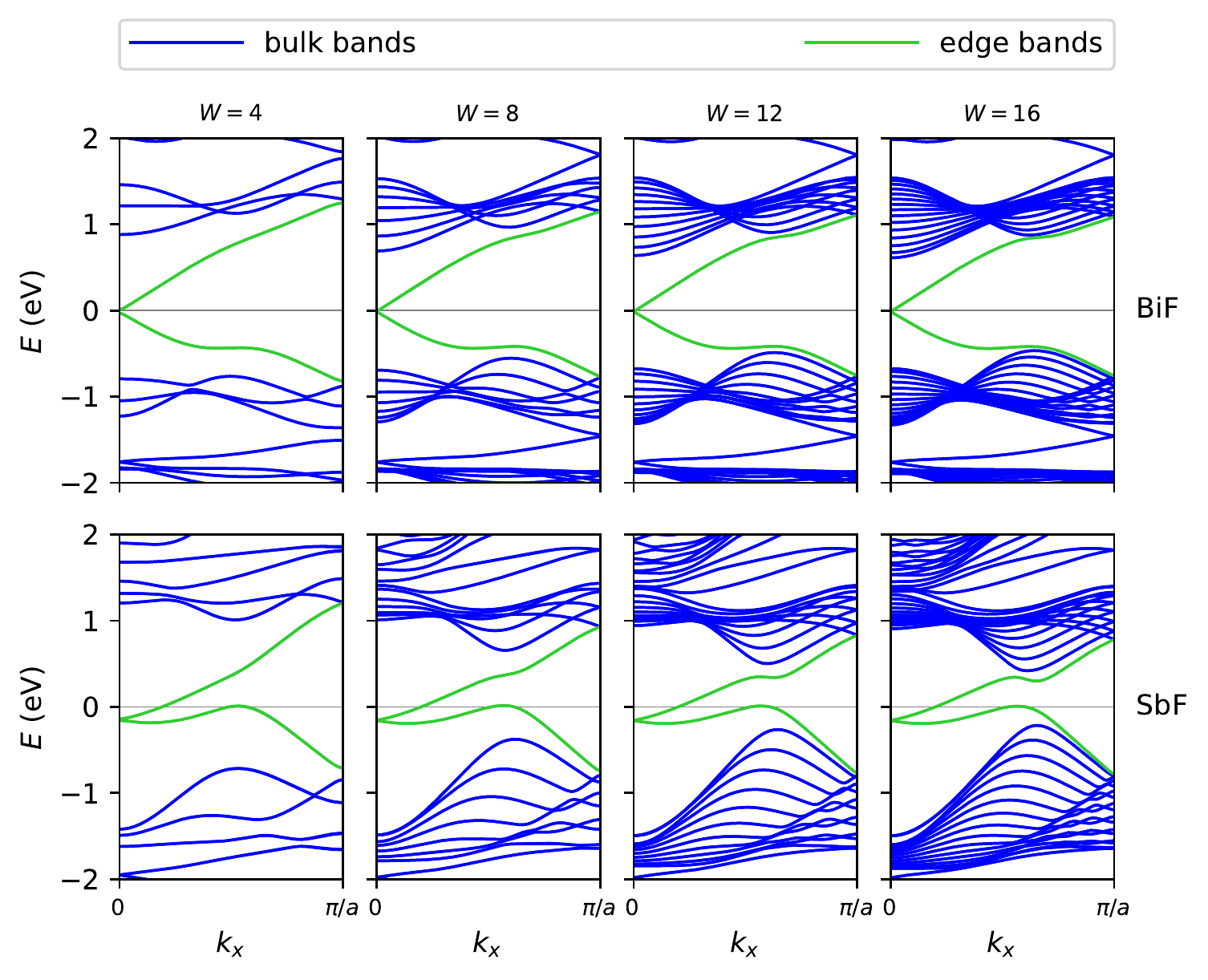}
	\includegraphics[width=0.49\linewidth]{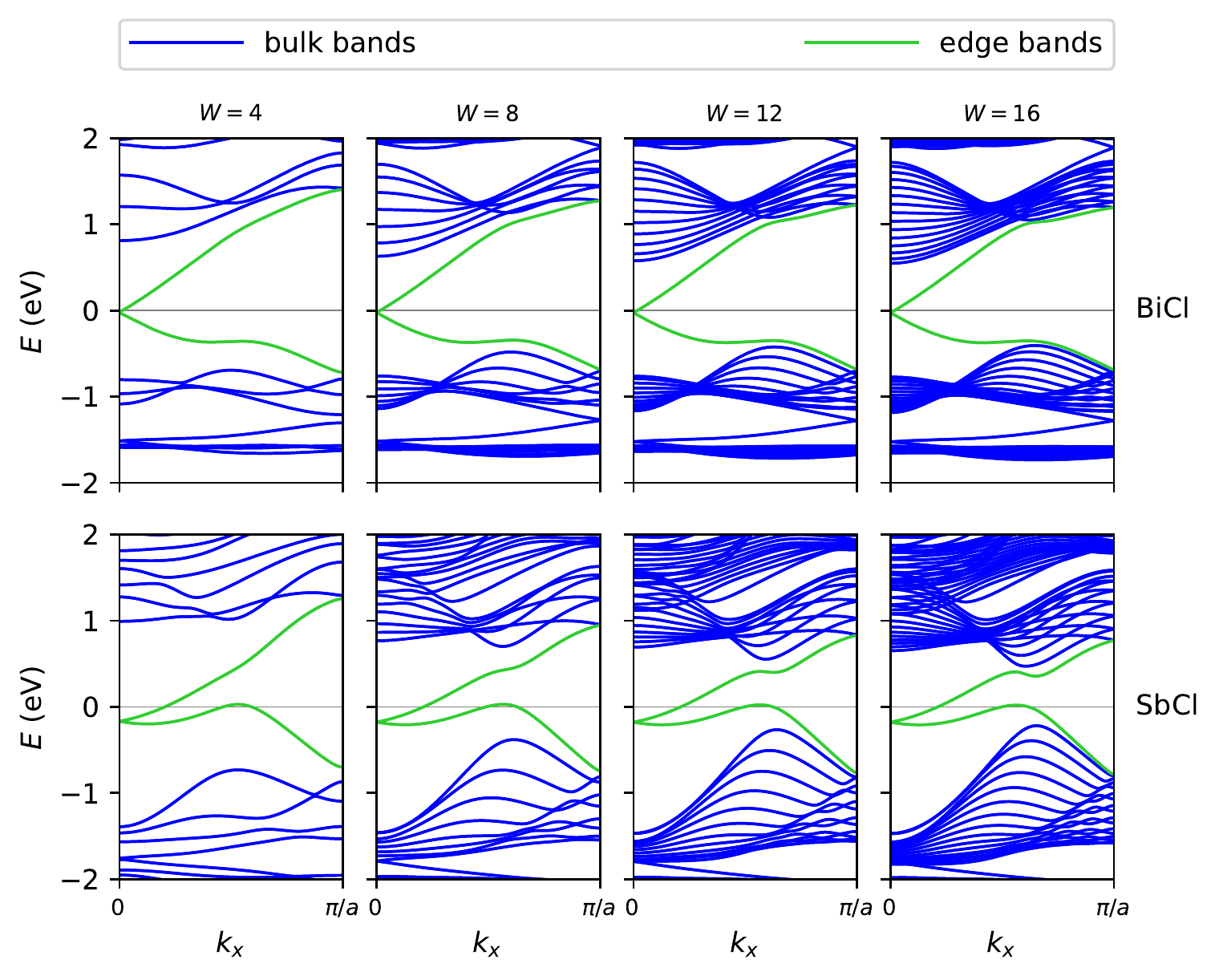}
	\includegraphics[width=0.49\linewidth]{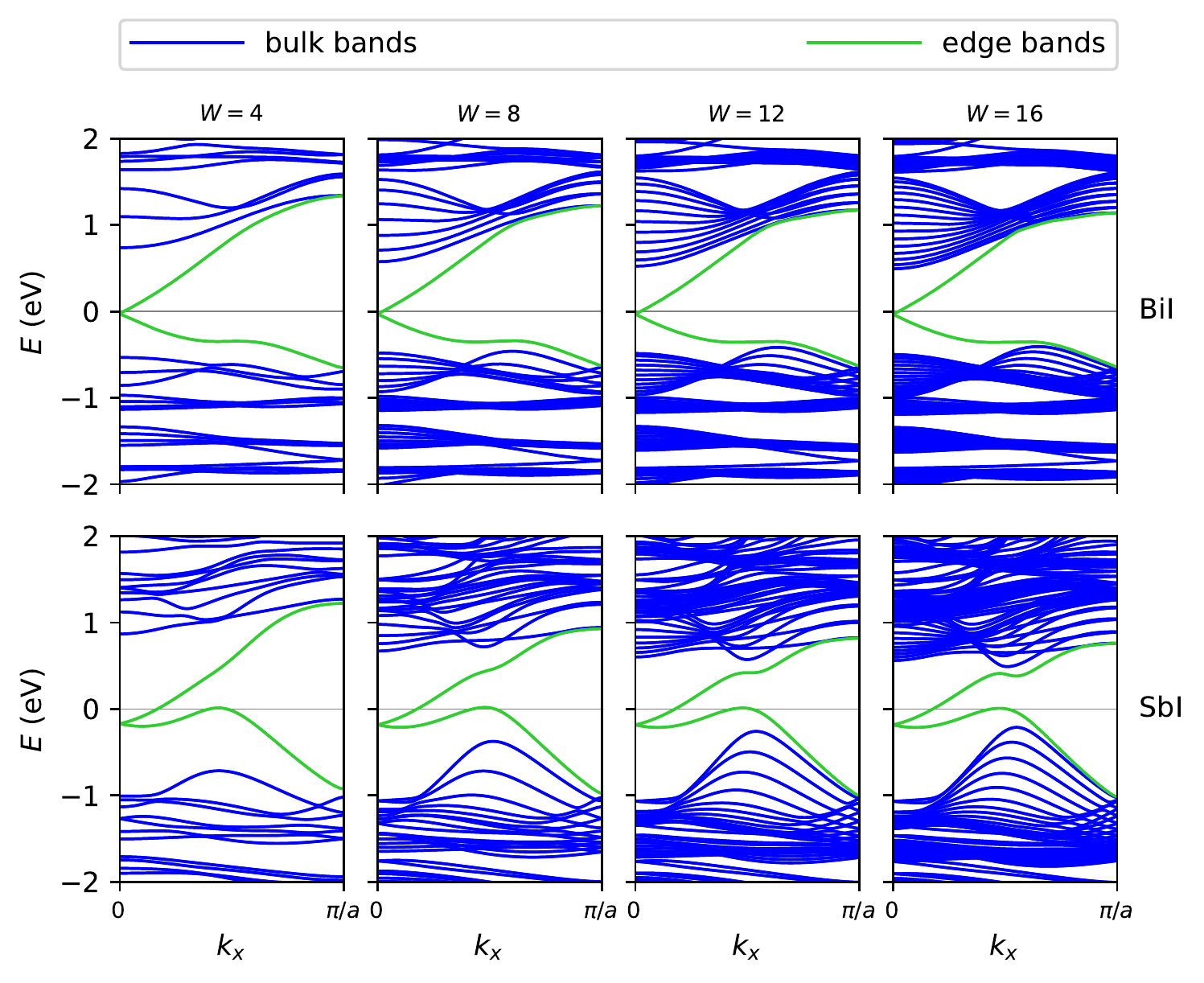}
	\caption{Band structure of BiX and SbX zigzag nanoribbons (X = F, Cl, I) of different width $W$. Topological edge states are reported in green. The energy scale is referred to the Fermi energy.}
	\label{fig:zig}
\end{figure}

\begin{figure}
	\centering
	\includegraphics[width=0.9\linewidth]{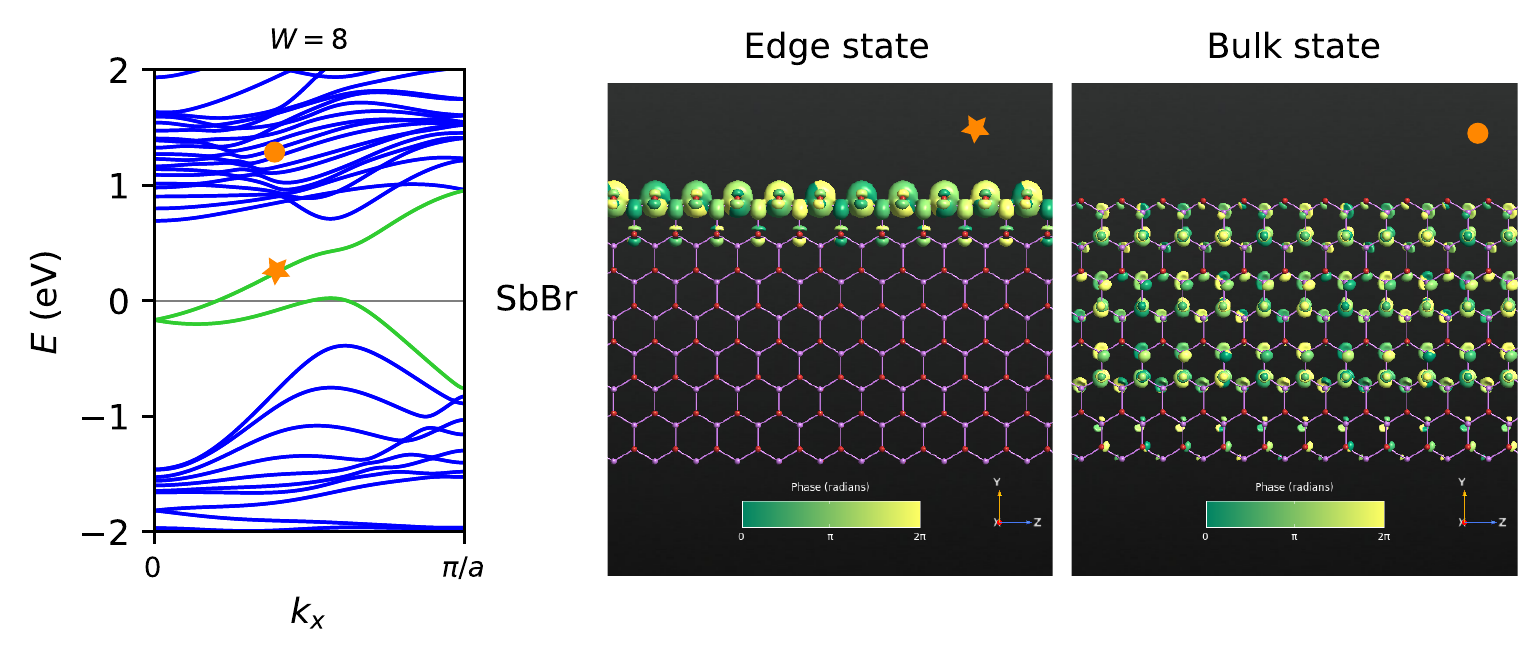}
	\caption{Wavefunctions of an SbBr nanoribbon of width $W=8$ for the electronic states indicated in the left panel. The figure shows the surface of constant wavefunction amplitude $|\psi(x,y,z)| = 0.05 \mathrm{\AA}^{-3/2}$, while the color scale indicates the phase.}
	\label{fig:wavefunction}
\end{figure}

The fact that such states are localized along the edge is confirmed by looking at their associated wavefunction, which we show in Fig.\ \ref{fig:wavefunction}. The figure clearly shows that Bloch eigenstates for the in-gap states are sharply localized near the edge of the nanoribbon, while electronic states deep into the conduction band are spread over the entire width of the ribbon.

\FloatBarrier
\subsection{Armchair nanoribbons}

\begin{figure}
	\centering
	\includegraphics[width=0.6\linewidth]{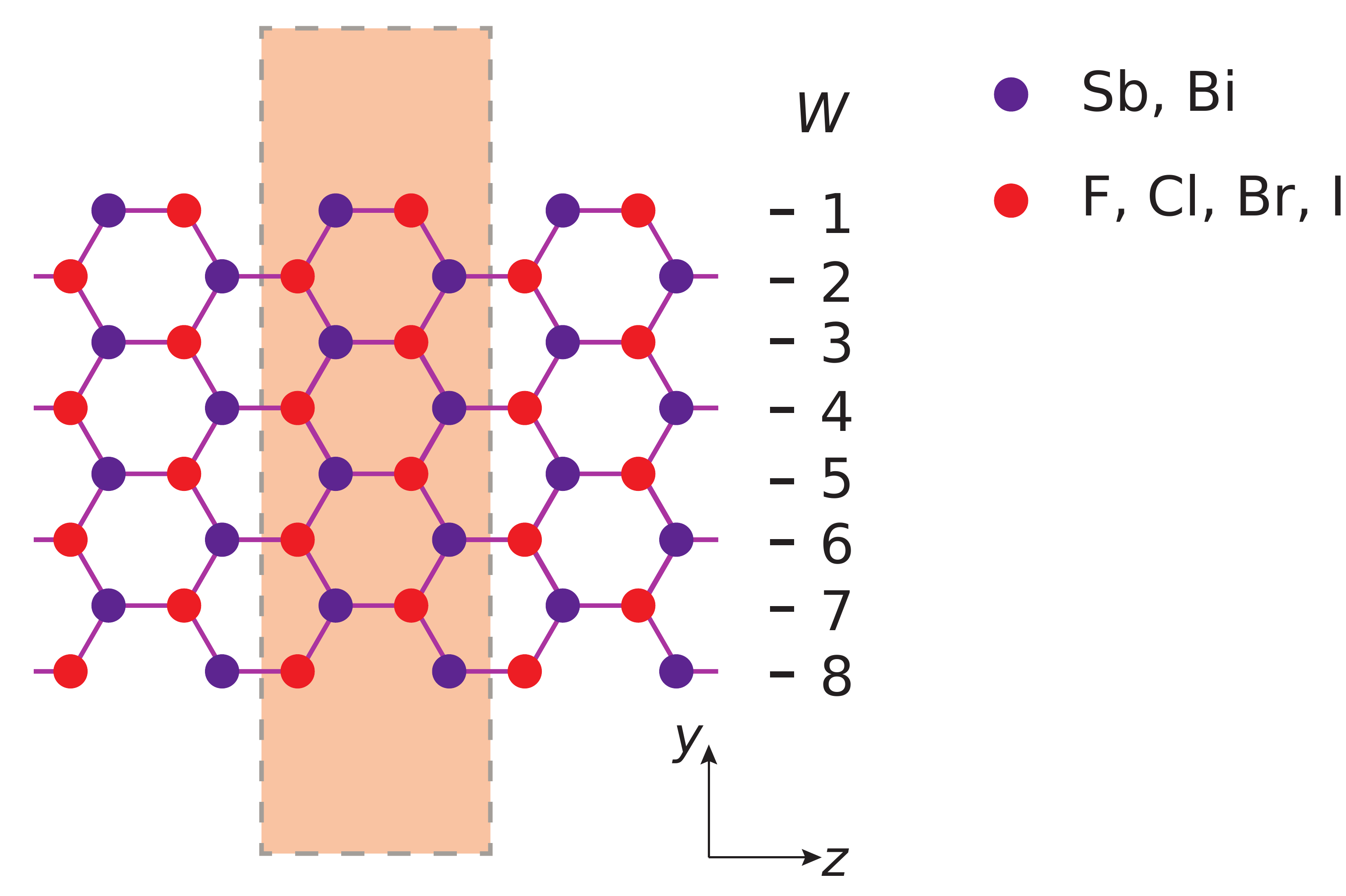}
	\caption{Top view of a Bismuth or Antimony halide armchair nanoribbon of width $W=8$. Dark purple dots denote the positions of Bi or Sb atoms, while halogen atoms (F, Cl, Br or I) are denoted in red. The shaded region corresponds to the unit cell.}
	\label{fig:ribbon_armchair}
\end{figure}

\begin{figure}
	\centering
	\includegraphics[width=0.49\linewidth]{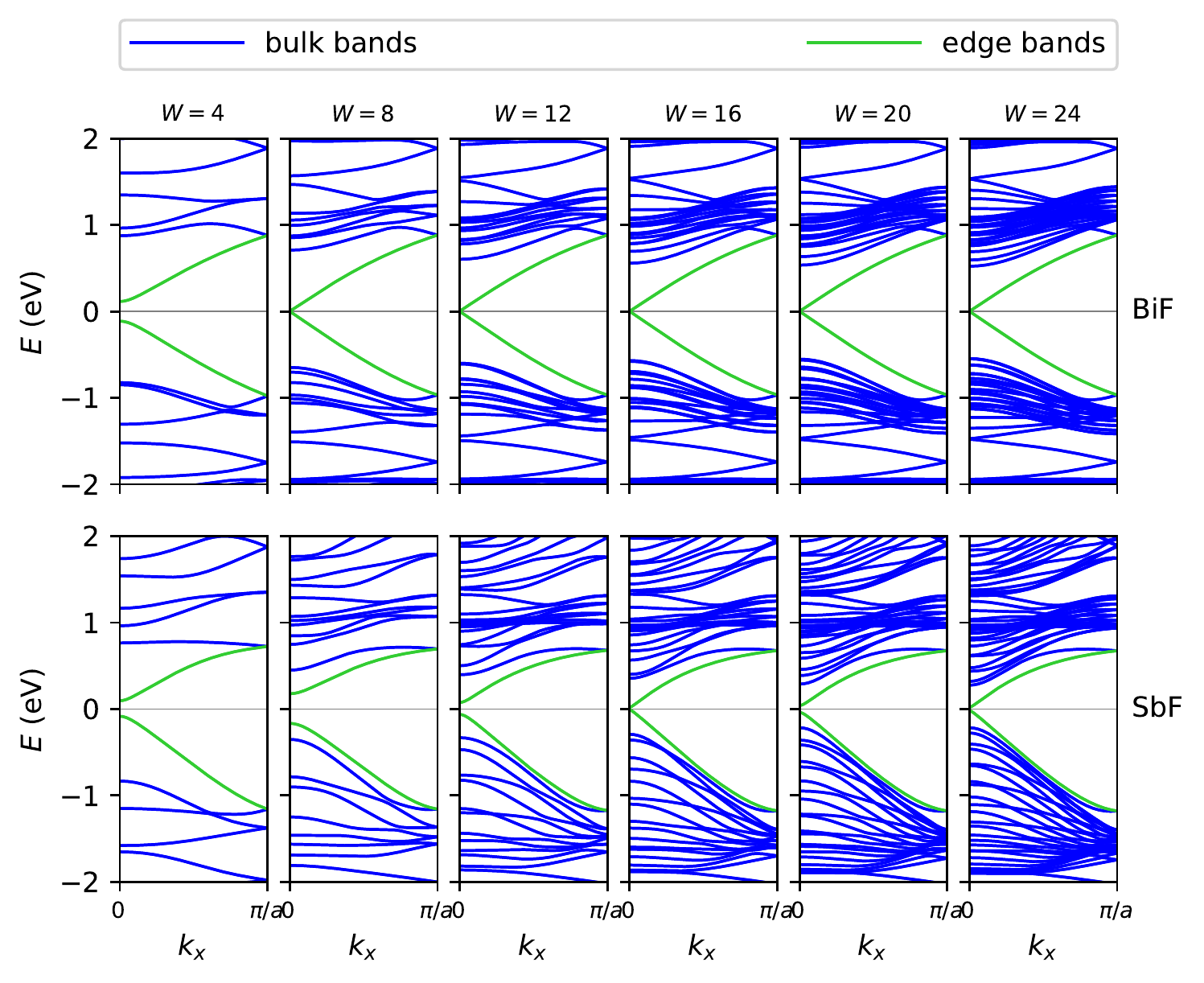}
	\includegraphics[width=0.49\linewidth]{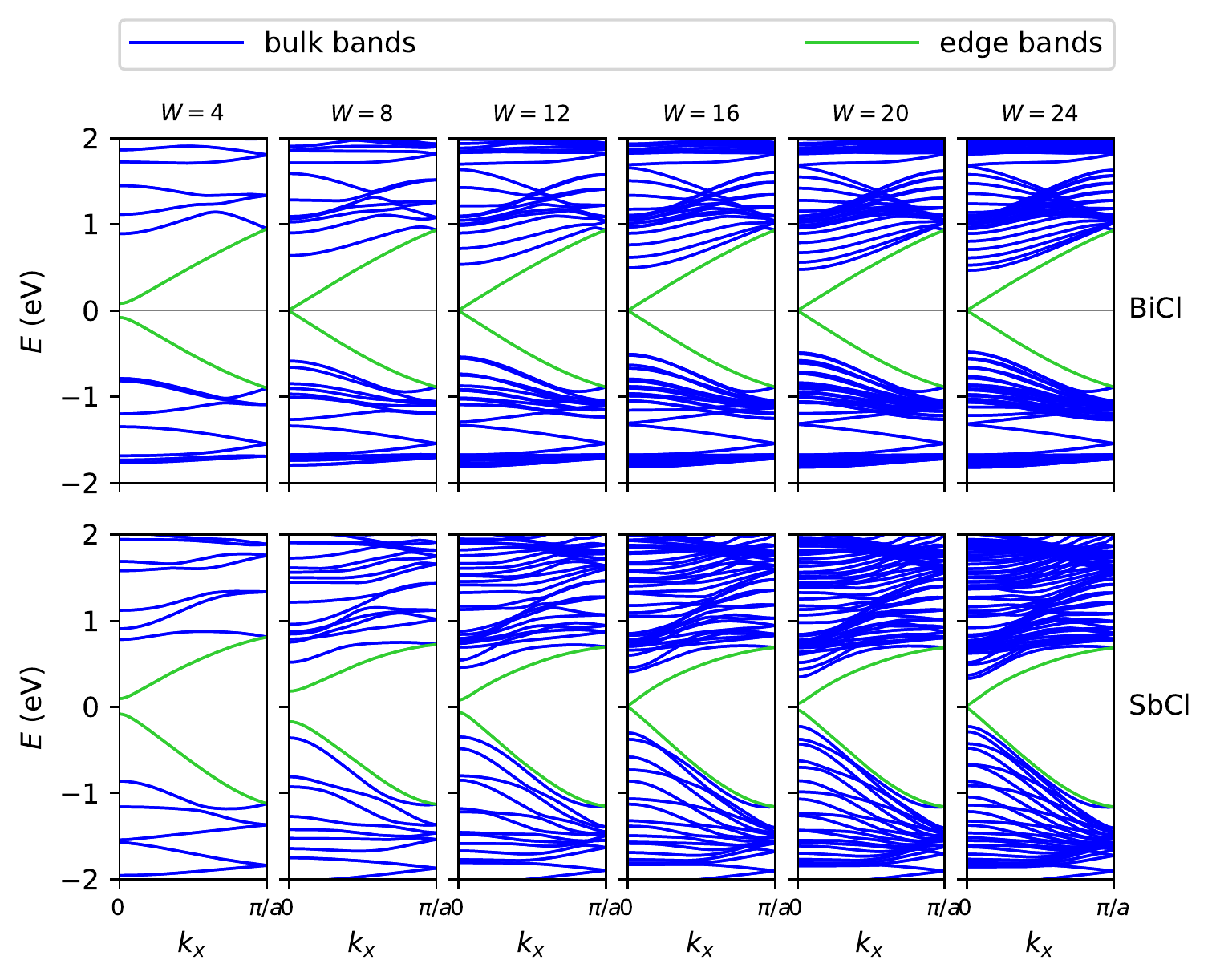}
	\includegraphics[width=0.49\linewidth]{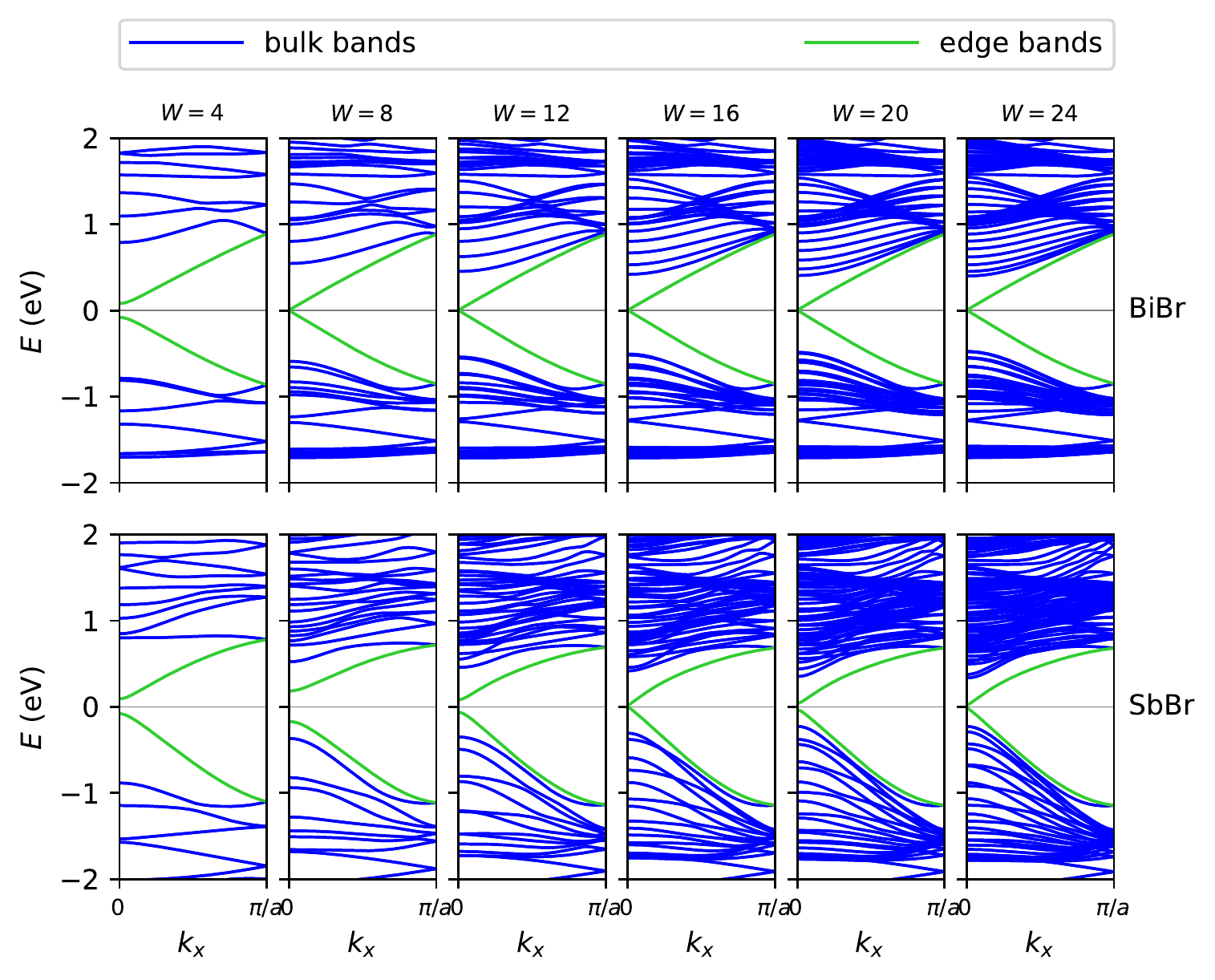}
	\includegraphics[width=0.49\linewidth]{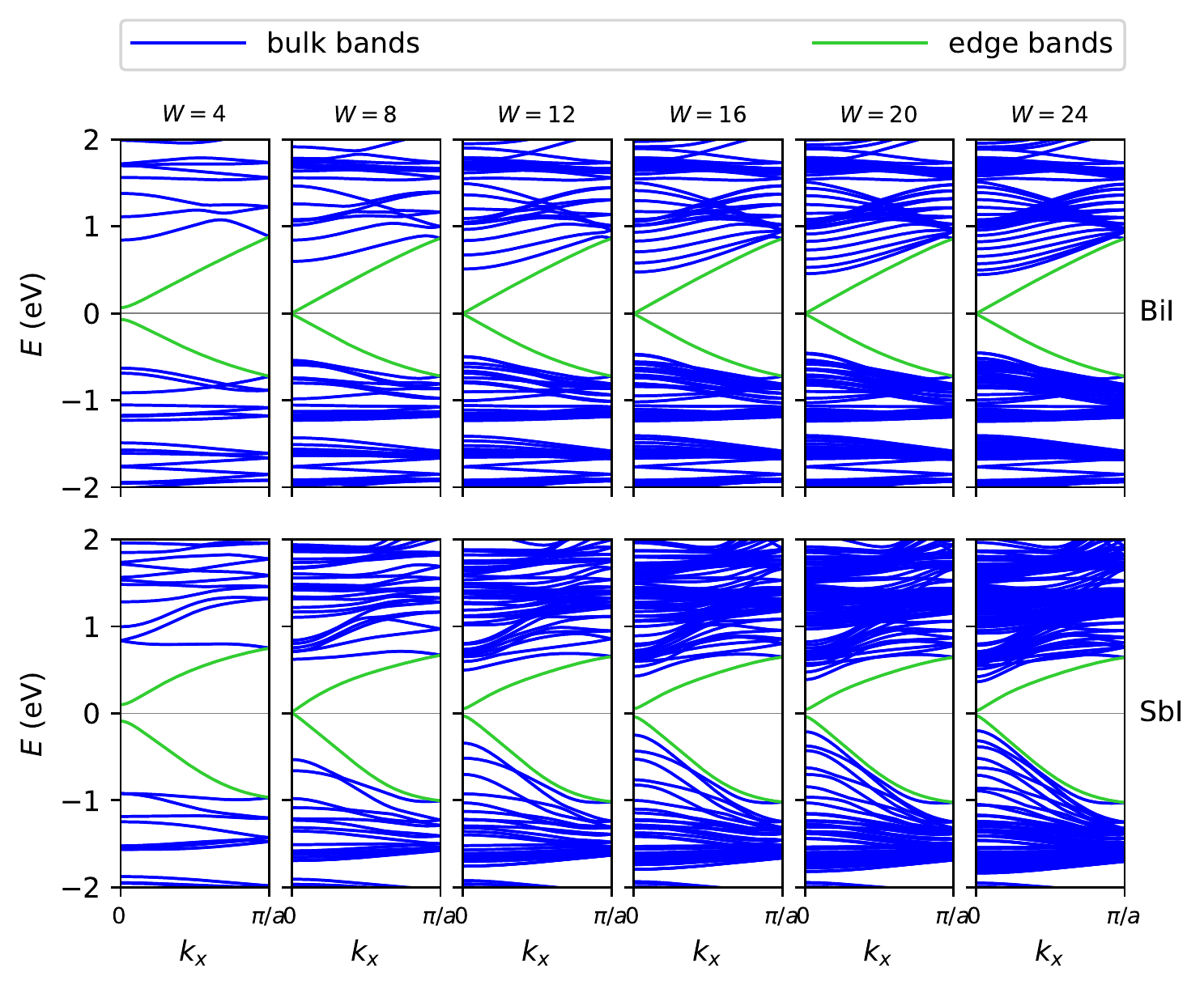}
	\caption{Band structure of BiX and SbX armchair nanoribbons (X = F, Cl, Br, I) of different width $W$. Topological edge states are reported in green. The energy scale is referred to the Fermi energy.}
	\label{fig:arm}
\end{figure}

In this section we show the electronic band structure for armchair nanoribbons of different width, whose crystal structure is depicted in Fig.\ \ref{fig:ribbon_armchair}.
As reported in Figure \ref{fig:arm}, all structures show the presence of additional metallic edge states with respect to the infinite bulk monolayer.
Note that the spectrum is still gapped for small armchair nanoribbons ($W=4-12$). This is due to hybridization between wavefunctions on opposite edges.
Indeed, in our notation, armchair nanoribbons of width $W$ are actually narrower by a factor $\sqrt{3}$ with respect to zigzag nanoribbon with identical $W$.
However, it is still interesting to note that the magnitude of the gap does not seem to decrease linearly with increasing $W$, especially for Antimony-based materials.

\FloatBarrier
\section{Transmission spectra for large ($W=8$) and narrow ($W=4$) ribbons}
\label{subsec:TS}

\subsection{Topological insulator nanoribbons}

\begin{figure}
	\centering
	\includegraphics[width=0.49\linewidth]{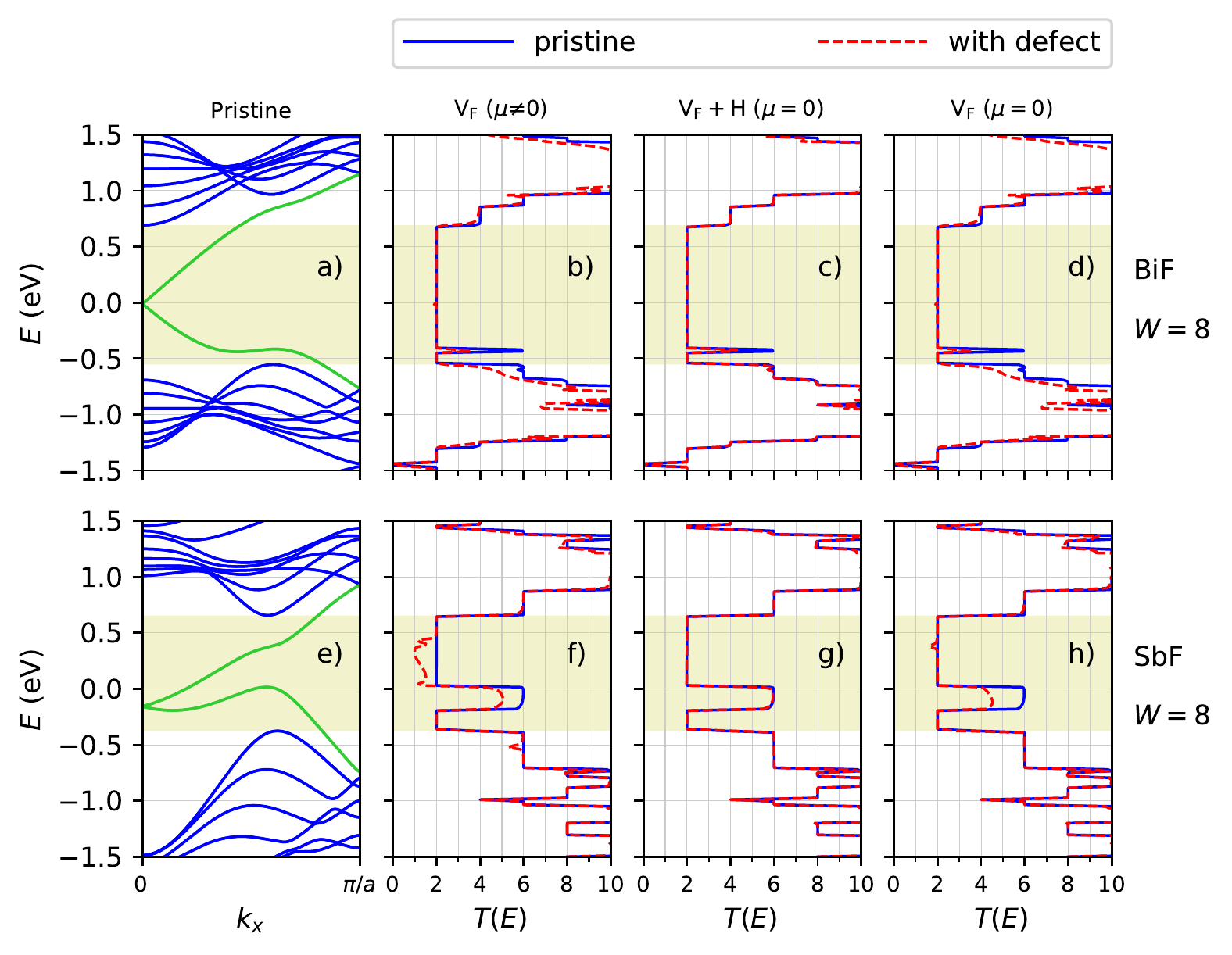}
	\includegraphics[width=0.49\linewidth]{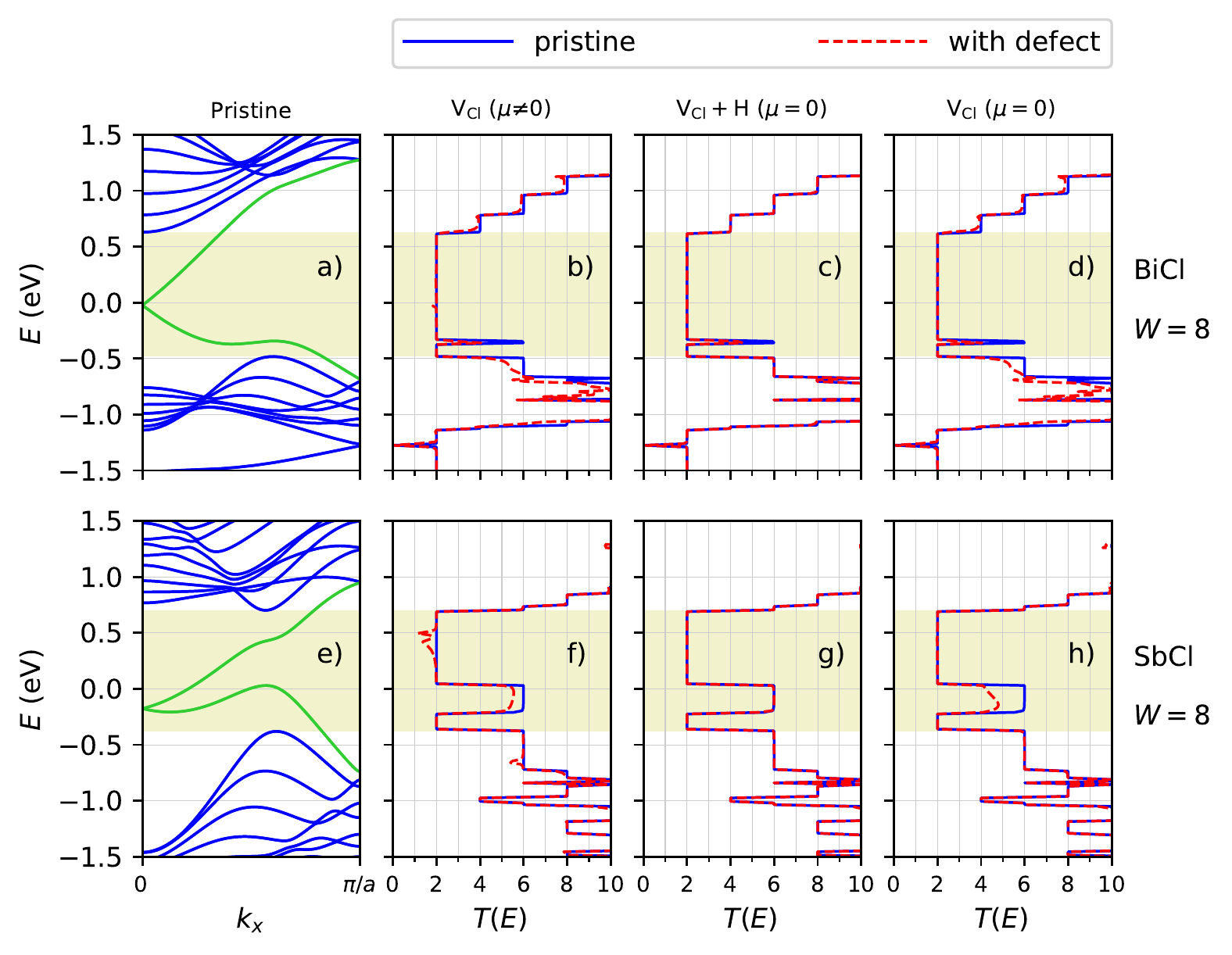}
	\includegraphics[width=0.49\linewidth]{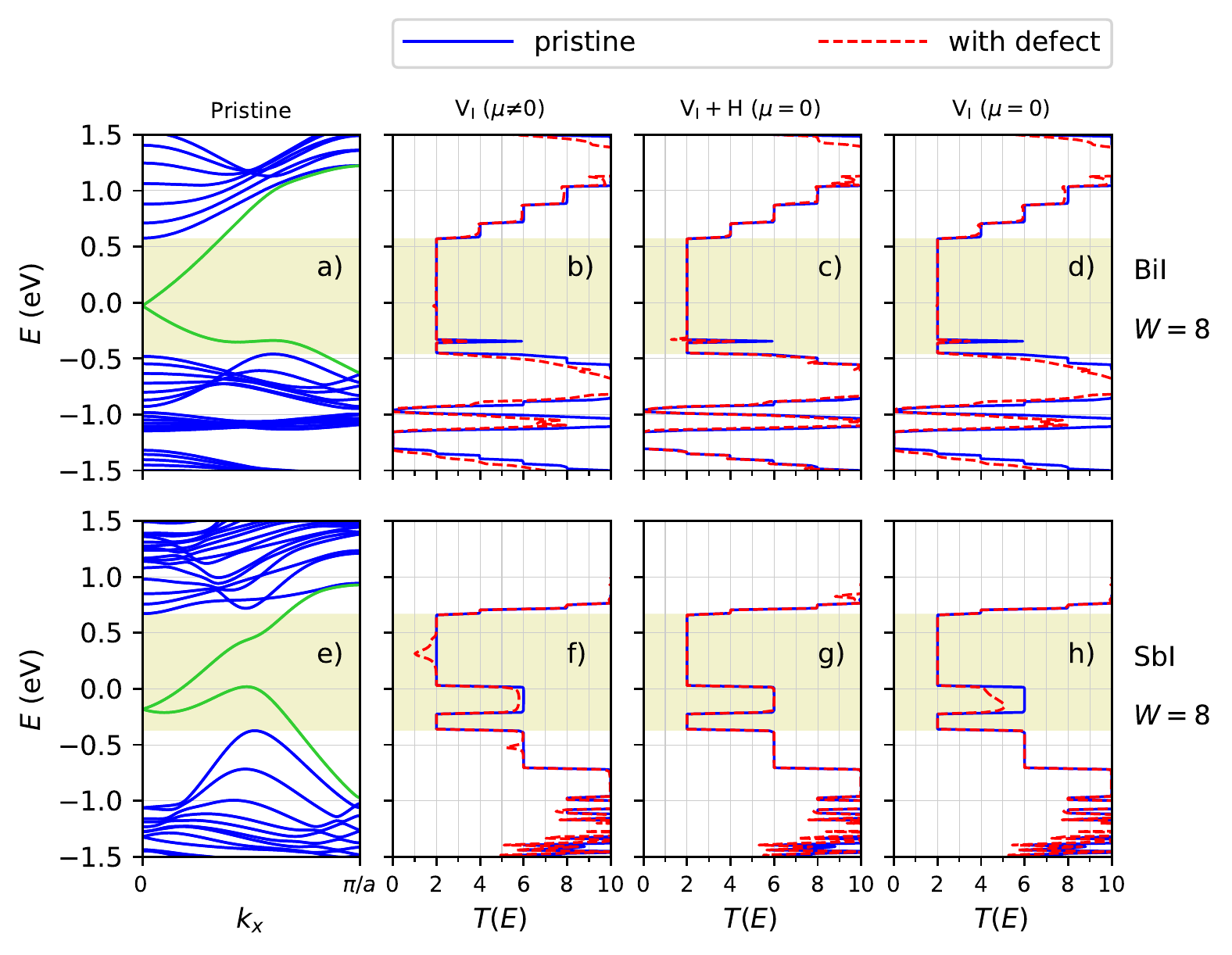}
	\caption{Transmission spectrum (TS) of BiX and SbX (X = F, Cl, I) zigzag nanoribbons of width $W=8$ in presence of one of the following edge defects: simple edge defect (b and f); Hydrogen-saturated edge defect (c and g); edge defect with zero magnetic moment (d and h). The TS for a pristine ribbon is also shown for comparison in each panel, and its band structure is reported in panels a and e.
	The energy region of insulating bulk is highlighted in yellow.}
	\label{fig:TS_edge-zig_8}
\end{figure}

\begin{figure}
	\centering
	\includegraphics[width=0.49\linewidth]{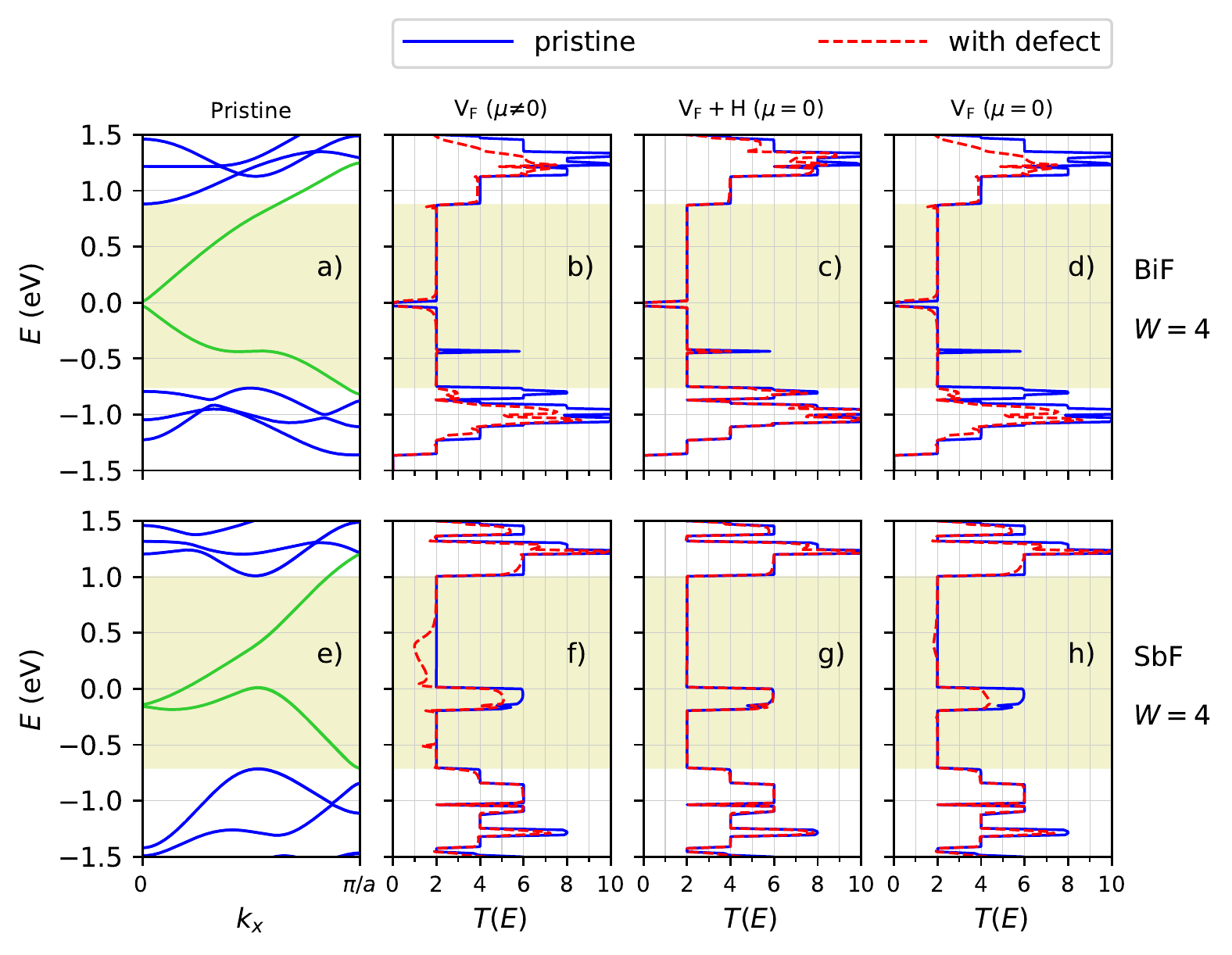}
	\includegraphics[width=0.49\linewidth]{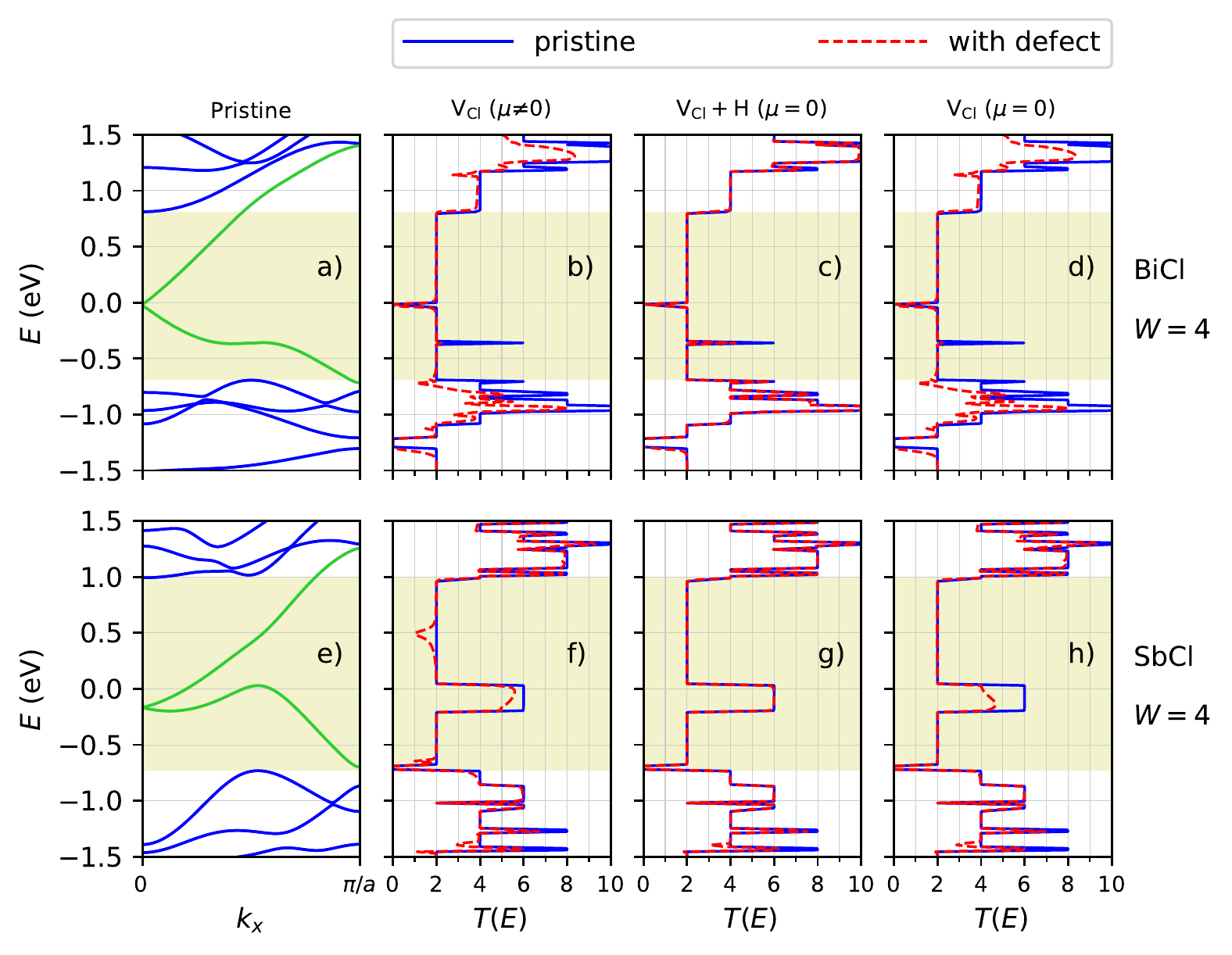}
	\includegraphics[width=0.49\linewidth]{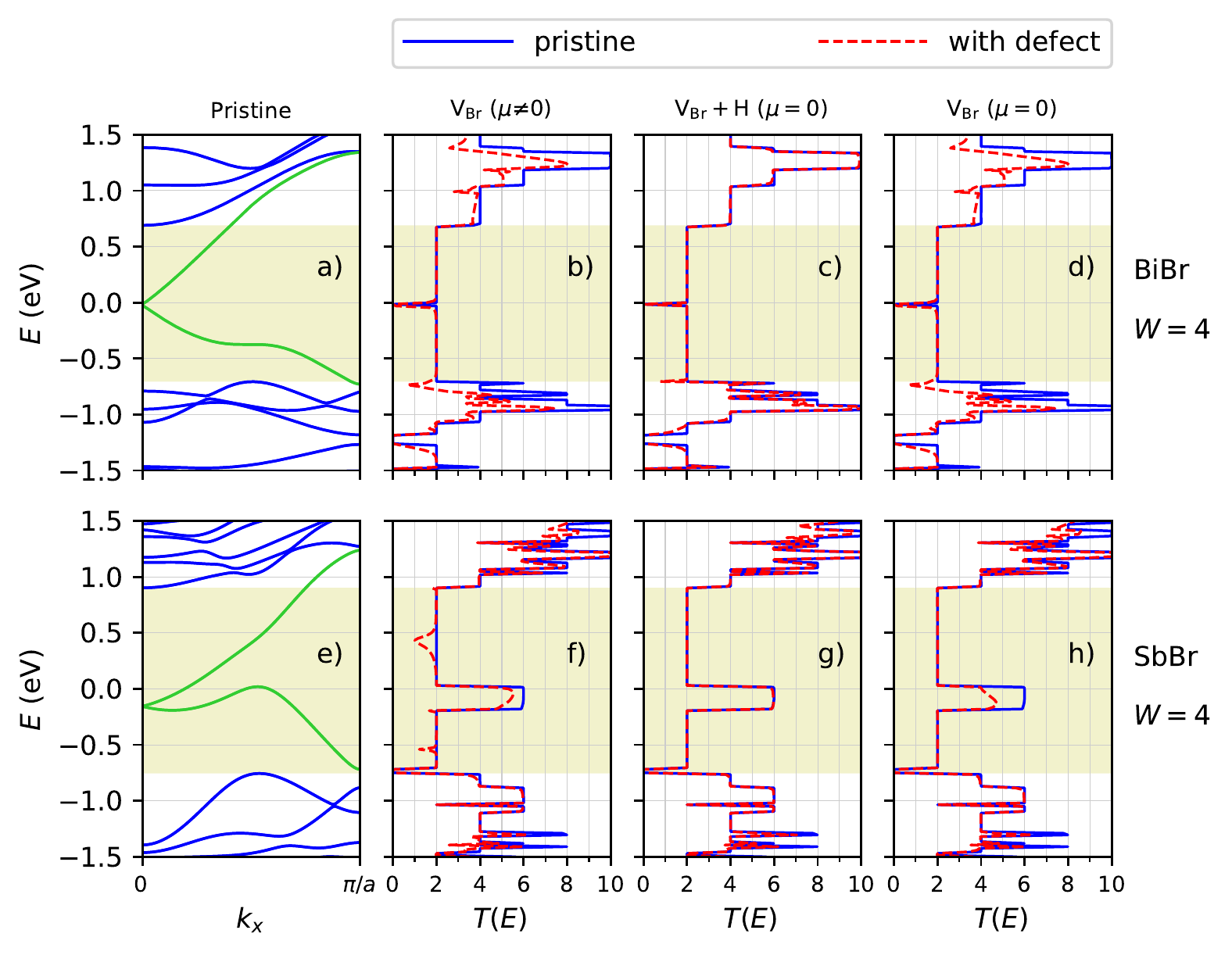}
	\includegraphics[width=0.49\linewidth]{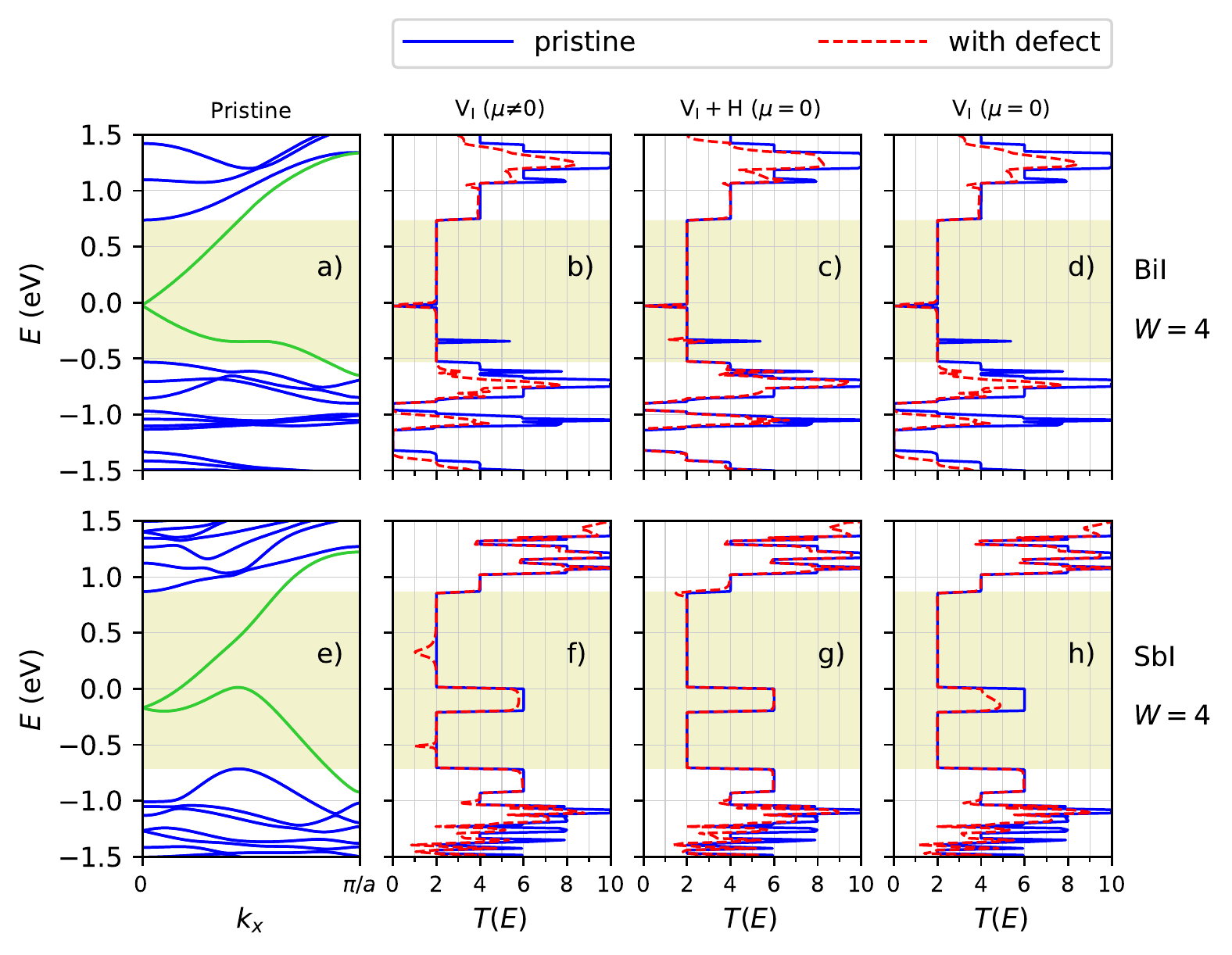}
	\caption{Transmission spectrum (TS) of all zigzag nanoribbons of width $W=4$ in presence of one of the following edge defects: simple edge defect (b and f); Hydrogen-saturated edge defect (c and g); edge defect with zero magnetic moment (d and h).
	The TS for a pristine ribbon is also shown for comparison in each panel, and its band structure is reported in panels a and e.
	The energy region of insulating bulk is highlighted in yellow.}
	\label{fig:TS_edge-zig_4}
\end{figure}

\begin{figure}
	\centering
	\includegraphics[width=0.2\linewidth]{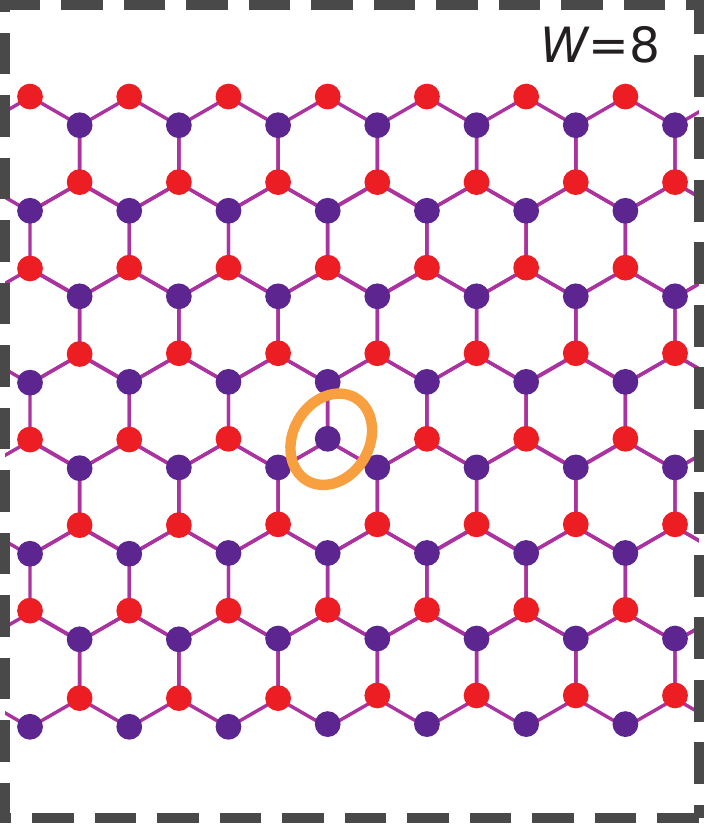}\\
	\vspace{5mm}
	\includegraphics[width=0.49\linewidth]{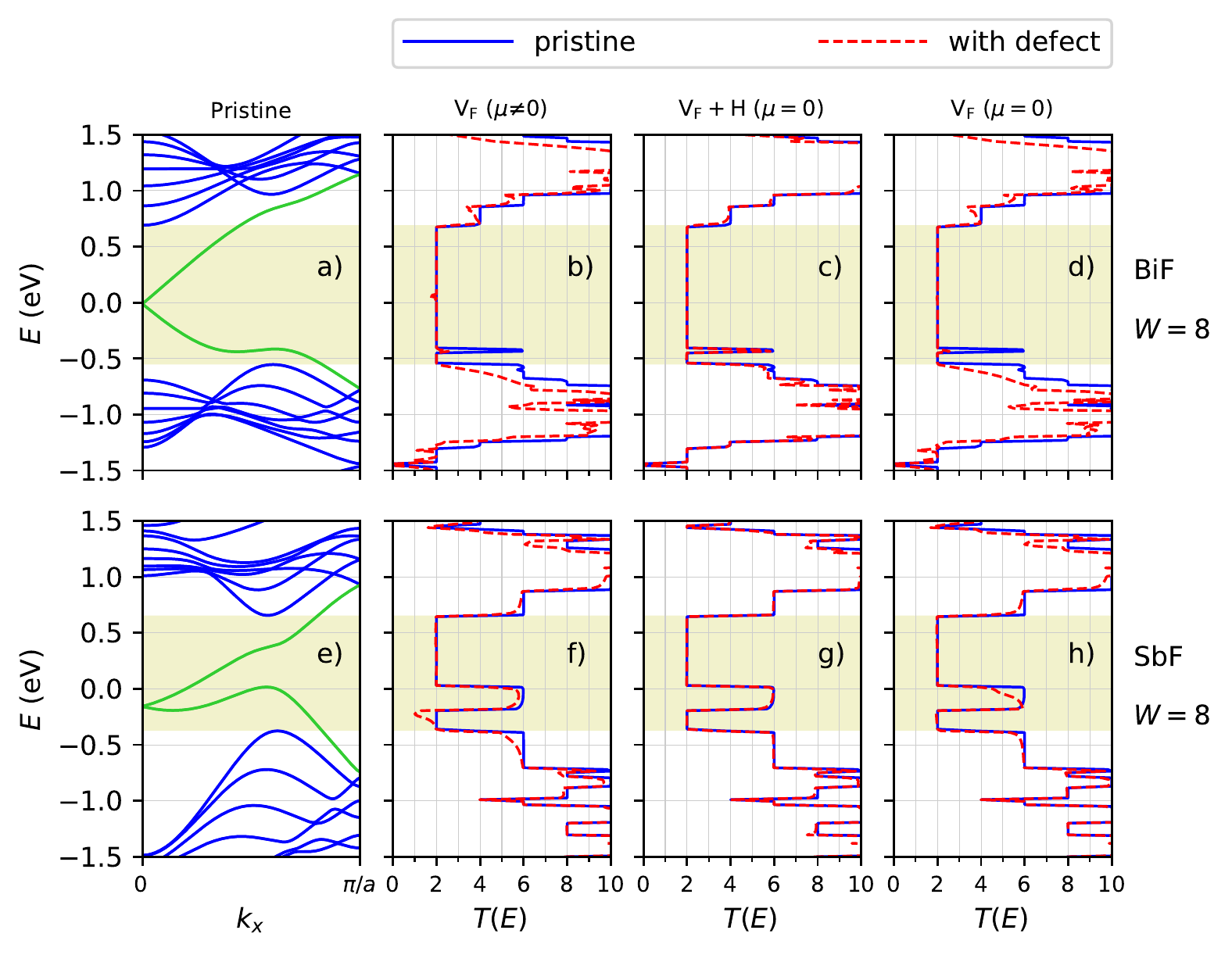}
	\includegraphics[width=0.49\linewidth]{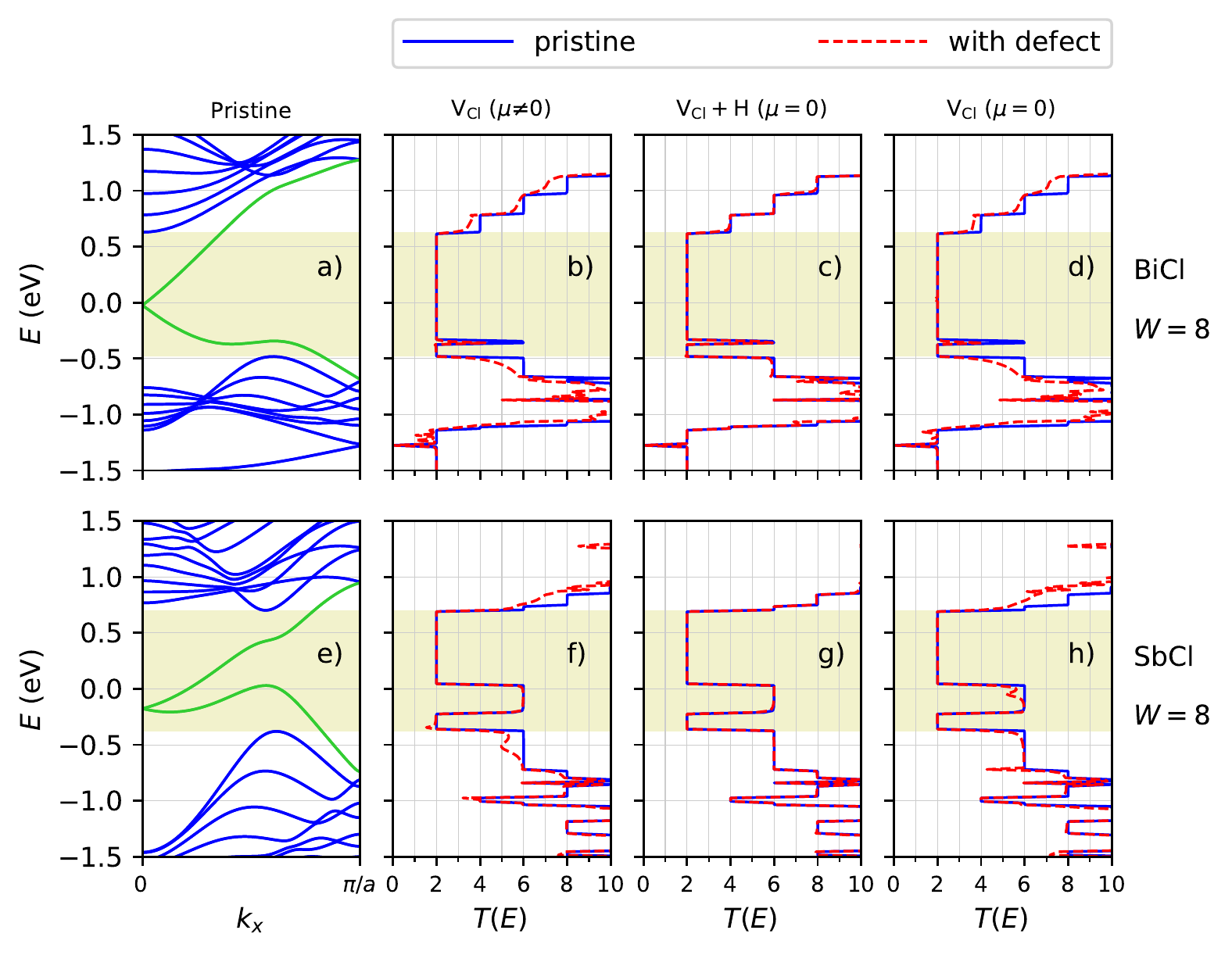}
	\includegraphics[width=0.49\linewidth]{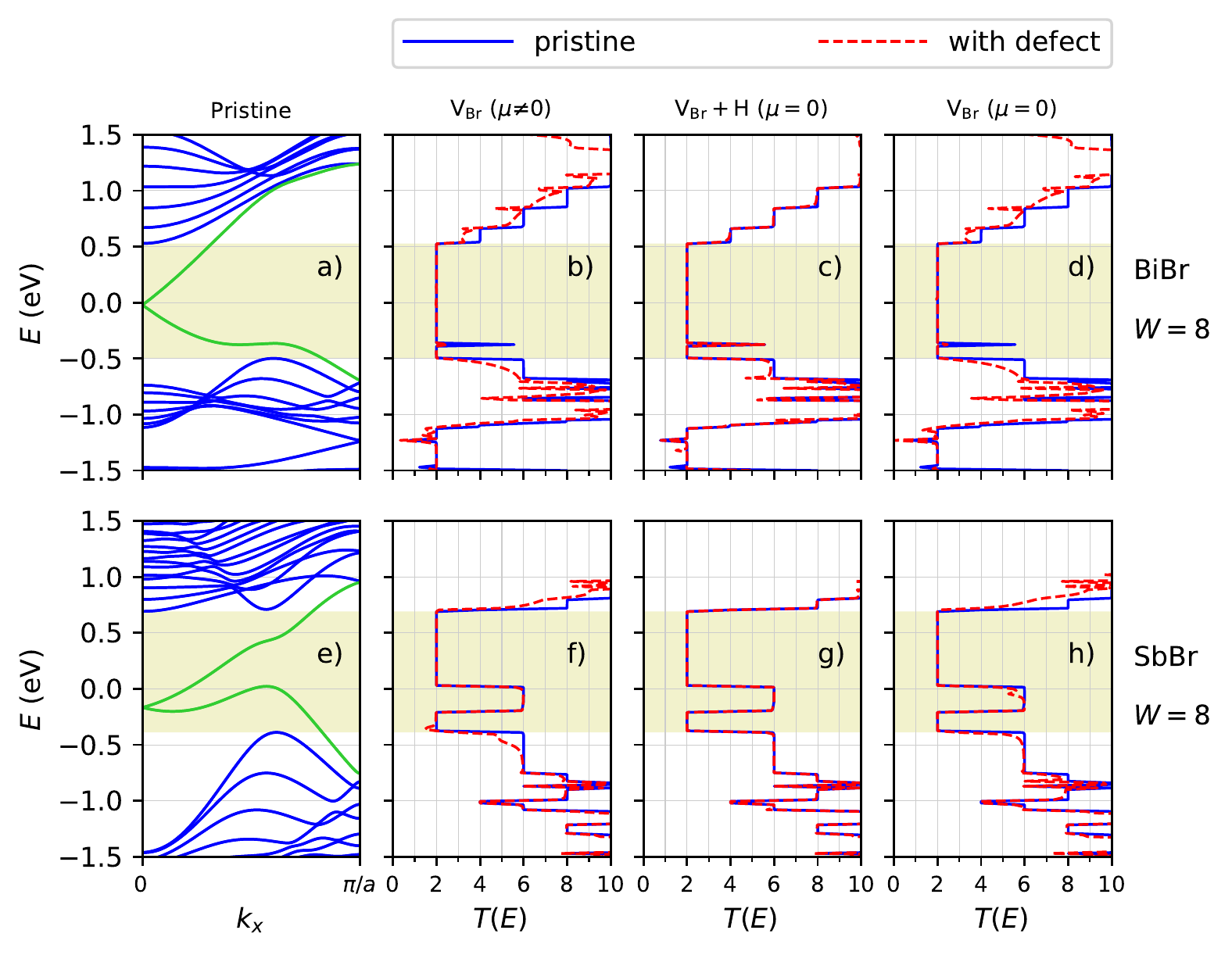}
	\includegraphics[width=0.49\linewidth]{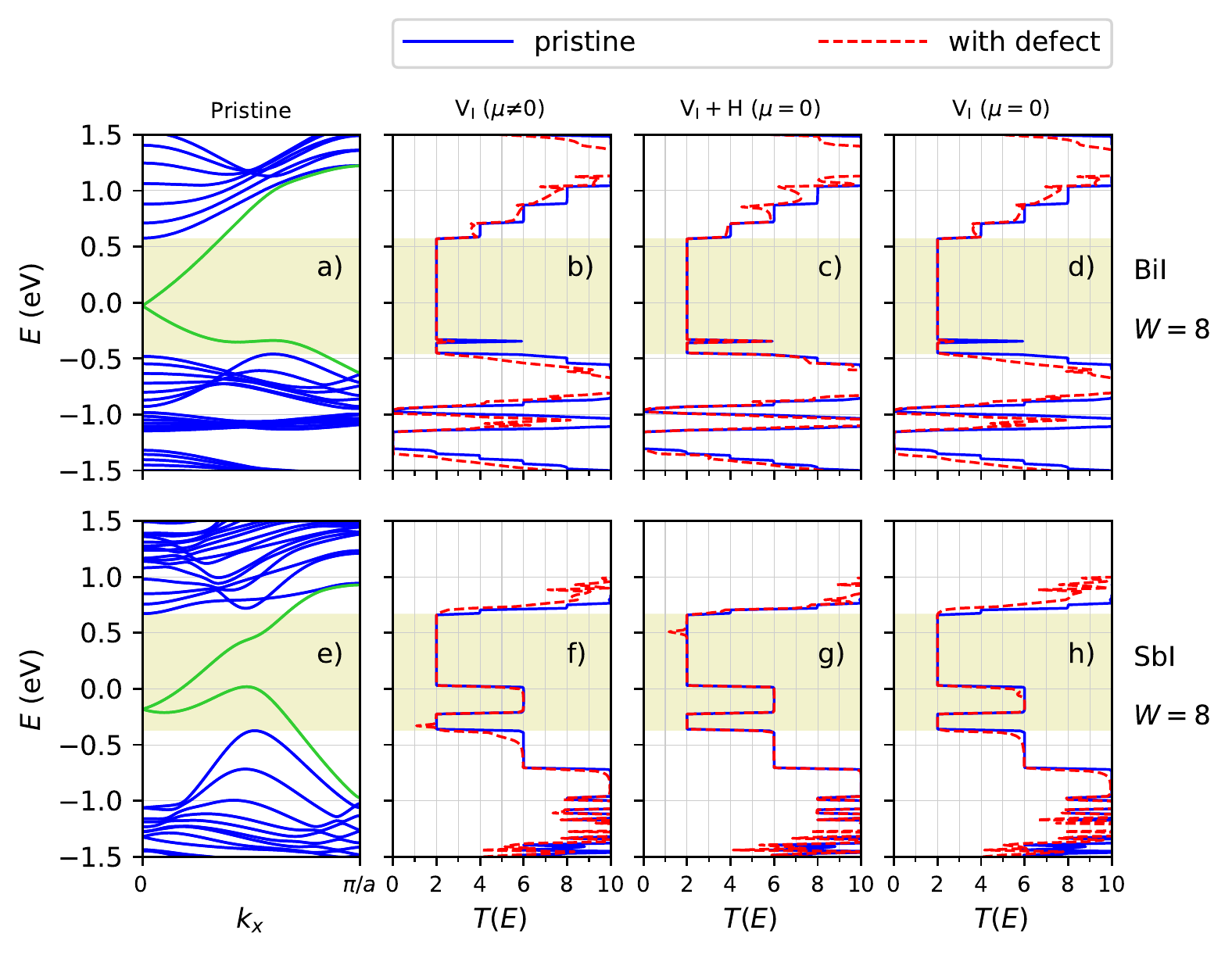}
	\caption{Transmission spectrum (TS) of all zigzag nanoribbons of width $W=8$ in presence of one of the following bulk defects: simple bulk defect (b and f); Hydrogen-saturated bulk defect (c and g); bulk defect with zero magnetic moment (d and h).
	The TS for a pristine ribbon is also shown for comparison in each panel, and its band structure is reported in panels a and d.
	The energy region of insulating bulk is highlighted in yellow.
	A sketch of the structure with the position of the defect is shown in the upper panel.}
	\label{fig:TS_bulk-zig_8}
\end{figure}

\begin{figure}
	\centering
	\includegraphics[width=0.2\linewidth]{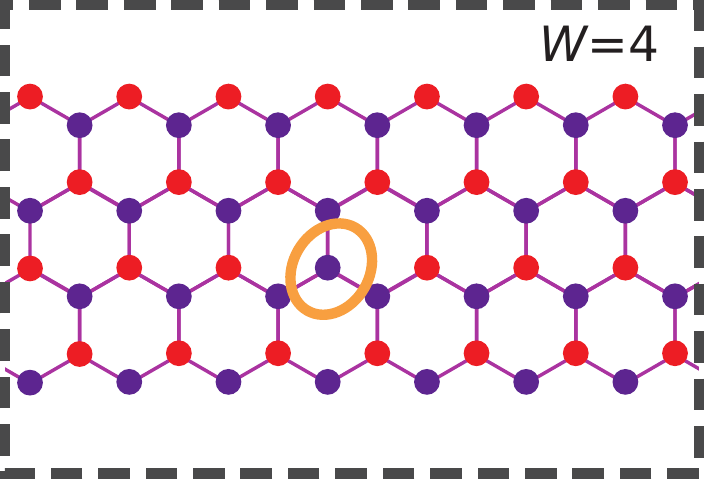}\\
	\vspace{5mm}
	\includegraphics[width=0.49\linewidth]{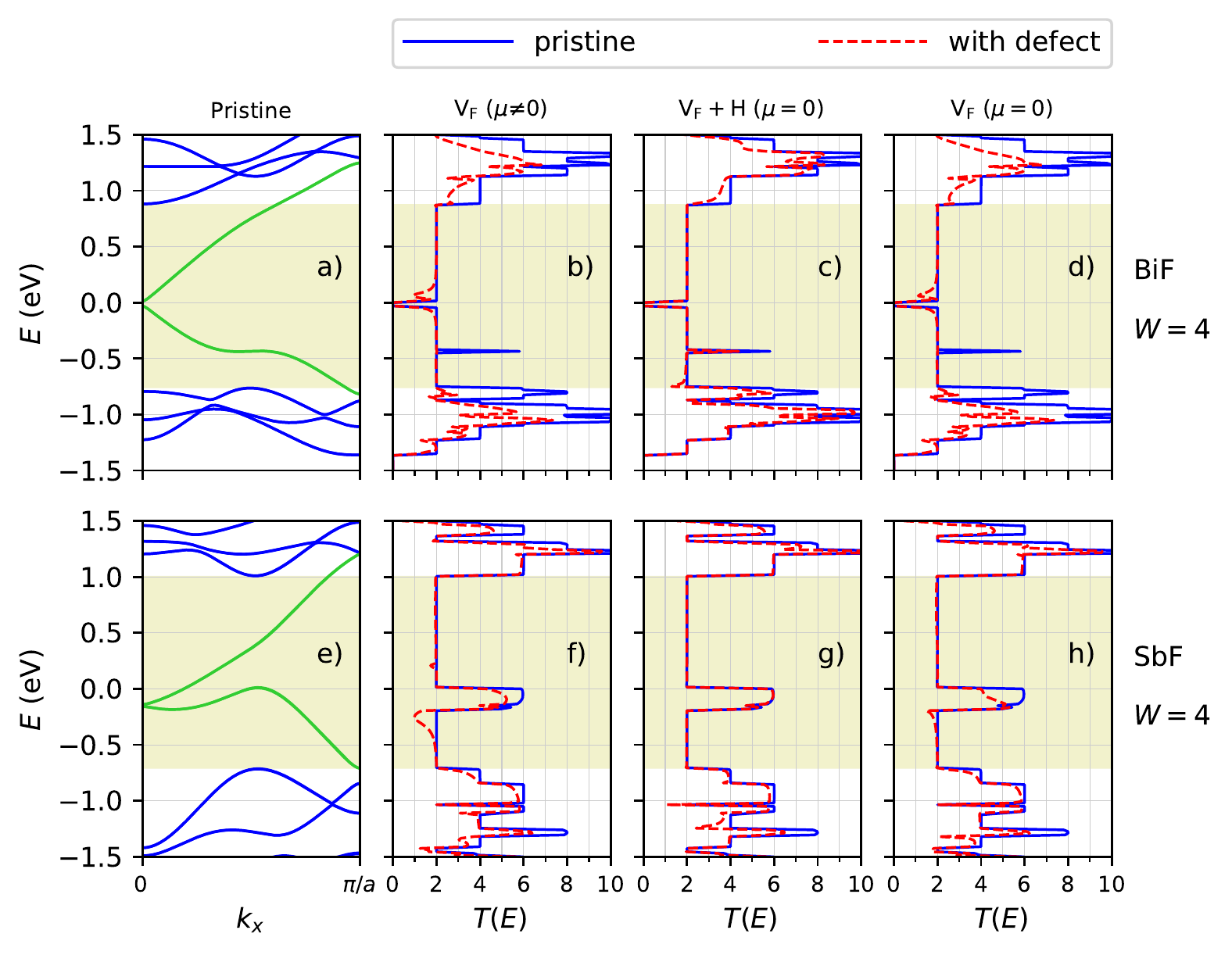}
	\includegraphics[width=0.49\linewidth]{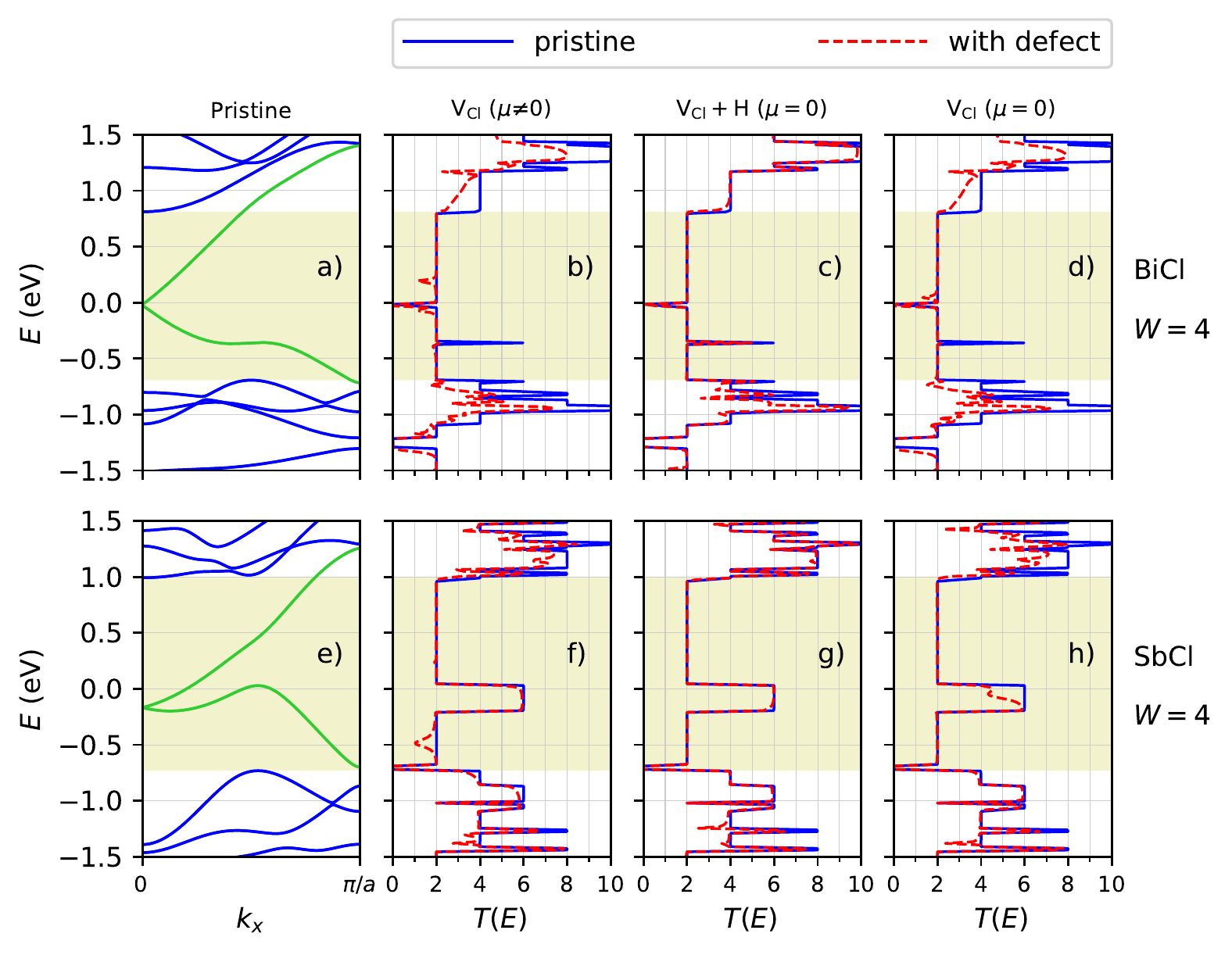}
	\includegraphics[width=0.49\linewidth]{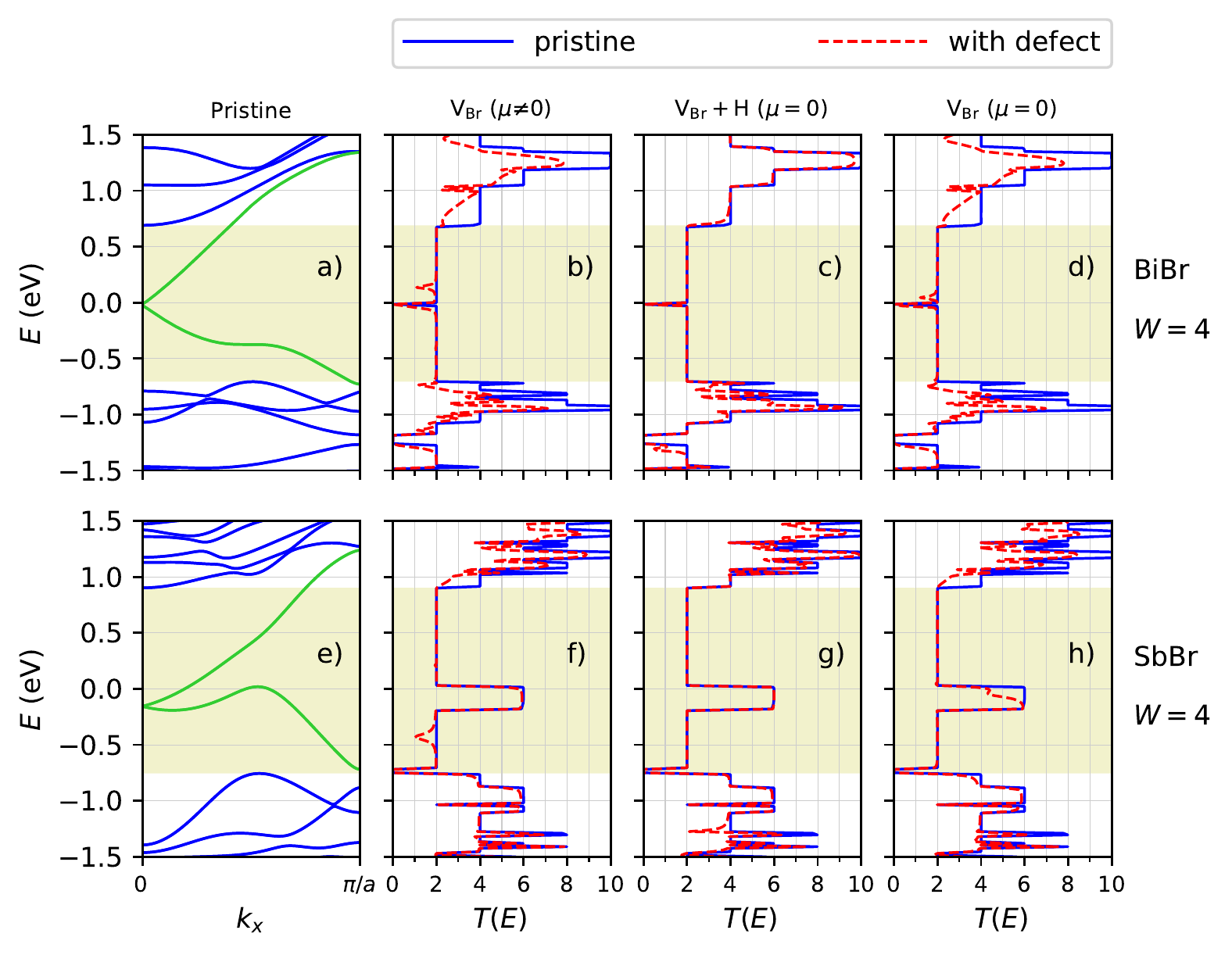}
	\includegraphics[width=0.49\linewidth]{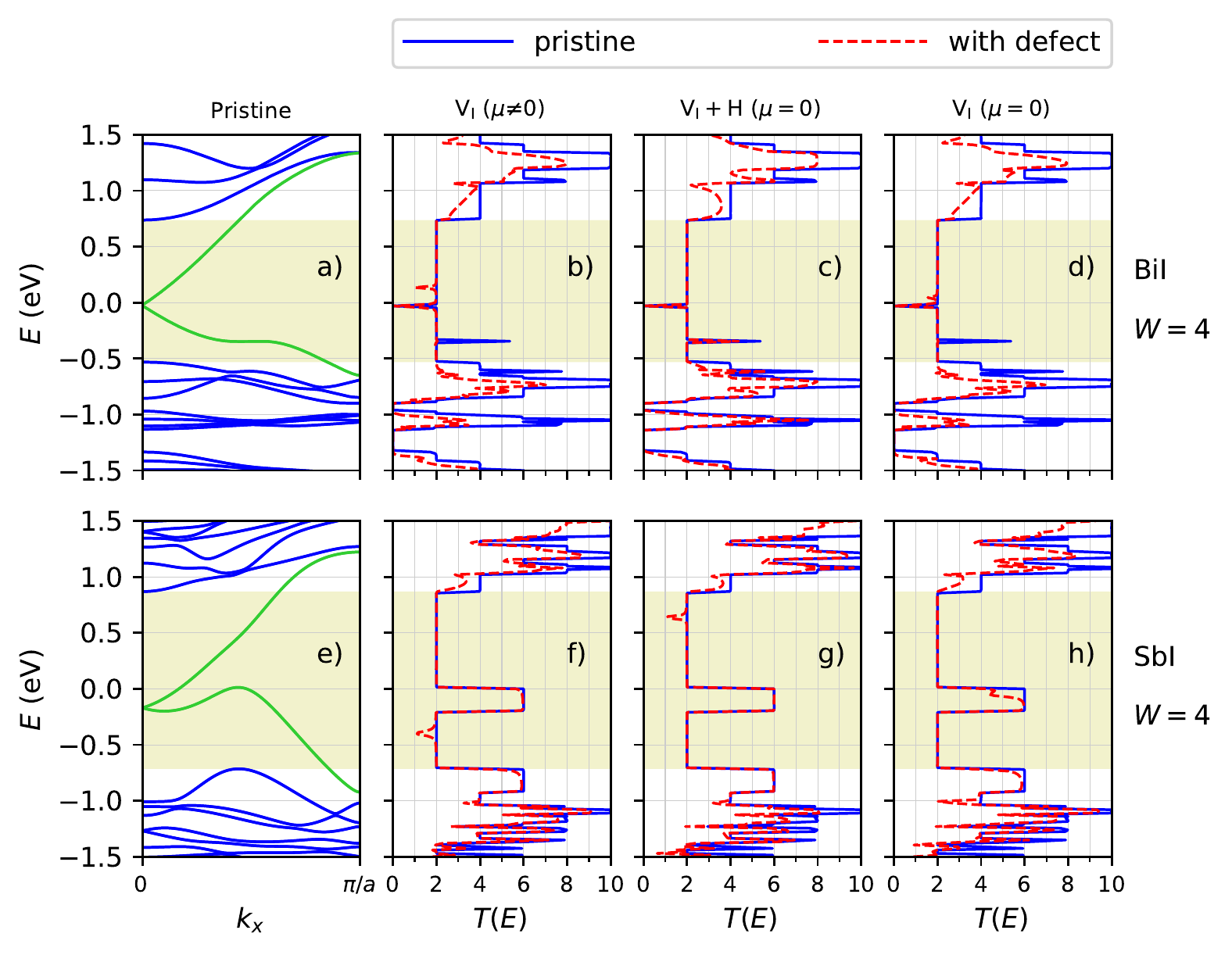}
	\caption{Transmission spectrum (TS) of all zigzag nanoribbons of width $W=4$ in presence of one of the following bulk defects: simple bulk defect (b and f); Hydrogen-saturated bulk defect (c and g); bulk defect with zero magnetic moment (d and h).
	The TS for a pristine ribbon is also shown for comparison in each panel, and its band structure is reported in panels a and d.
	The energy region of insulating bulk is highlighted in yellow.
	A sketch of the structure with the position of the defect is shown in the upper panel.}	
	\label{fig:TS_bulk-zig_4}
\end{figure}

In this section we report the transmission spectrum of all BiX and SbX zigzag nanoribbons of width $W=4$ and $W=8$ in the presence of either edge or bulk defects --- some of which have already been shown in the main text.
For all materials considered, we report three transmission spectra corresponding to the following three configurations:
\begin{itemize}
	\item $\mathrm{V_X},\ (\mu \ne 0)$: single edge/bulk defect with (possibly) non-vanishing magnetic moments. 
	
	\item $\mathrm{V_X + H},\ (\mu = 0)$: Hydrogen saturated edge/bulk defect, with no magnetic moment.
	
	\item $\mathrm{V_X},\ (\mu = 0)$: single edge/bulk defect with all magnetic moments set to zero. 
\end{itemize}
Figures\ \ref{fig:TS_edge-zig_8} and \ref{fig:TS_edge-zig_4} pertain to edge defects in nanoribbons of width $W=8$ and $W=4$ respectively. Similarly, Figs.\ \ref{fig:TS_bulk-zig_8} and \ref{fig:TS_bulk-zig_4} consider bulk defects in nanoribbons of width $W=8$ and $W=4$ respectively. 

Data shown here do not add new physics, but further demonstrate the phenomena already discussed in the main text for the case of BiBr and SbBr. In particular:
\begin{itemize}
	\item[(i)] SbX nanoribbons with edge defects show a partial suppression of the transmission in the form of a localized anti-resonance, due to the formation of a magnetic moment at the vacancy --- see panels f in Figs.\ \ref{fig:TS_edge-zig_8} and \ref{fig:TS_edge-zig_4} for nanoribbons of width $W=8$ and $W=4$ respectively. 
	
	\item[(ii)] Bulk defects do not affect transport for $W=8$, but lead to inter-edge scattering for $W=4$ which is driven by the partial overlap of the impurity state with the edge modes --- see Figs.\ \ref{fig:TS_bulk-zig_8} and \ref{fig:TS_bulk-zig_4}.
	
	\item[(iii)] For the case of Antimony-based materials, transport is not fully protected in the energy range where three pairs of edge states form at each interface. This is a manifestation of the underlying $\mathbb Z_2$ invariance, as we discuss in the main text. Note that the transmission spectrum never drops below 4.
	
	\item[(iv)] Chemical saturation of the dangling bonds with Hydrogen is generally sufficient to restore topological protection, since spurious magnetic moments are removed and the energy level of the defect are moved away from the bulk gap region.
\end{itemize}

\FloatBarrier
\subsection{Non-topological (MoS$_2$) nanoribbons}

\begin{figure}
	\centering
	\includegraphics[width=0.9\linewidth]{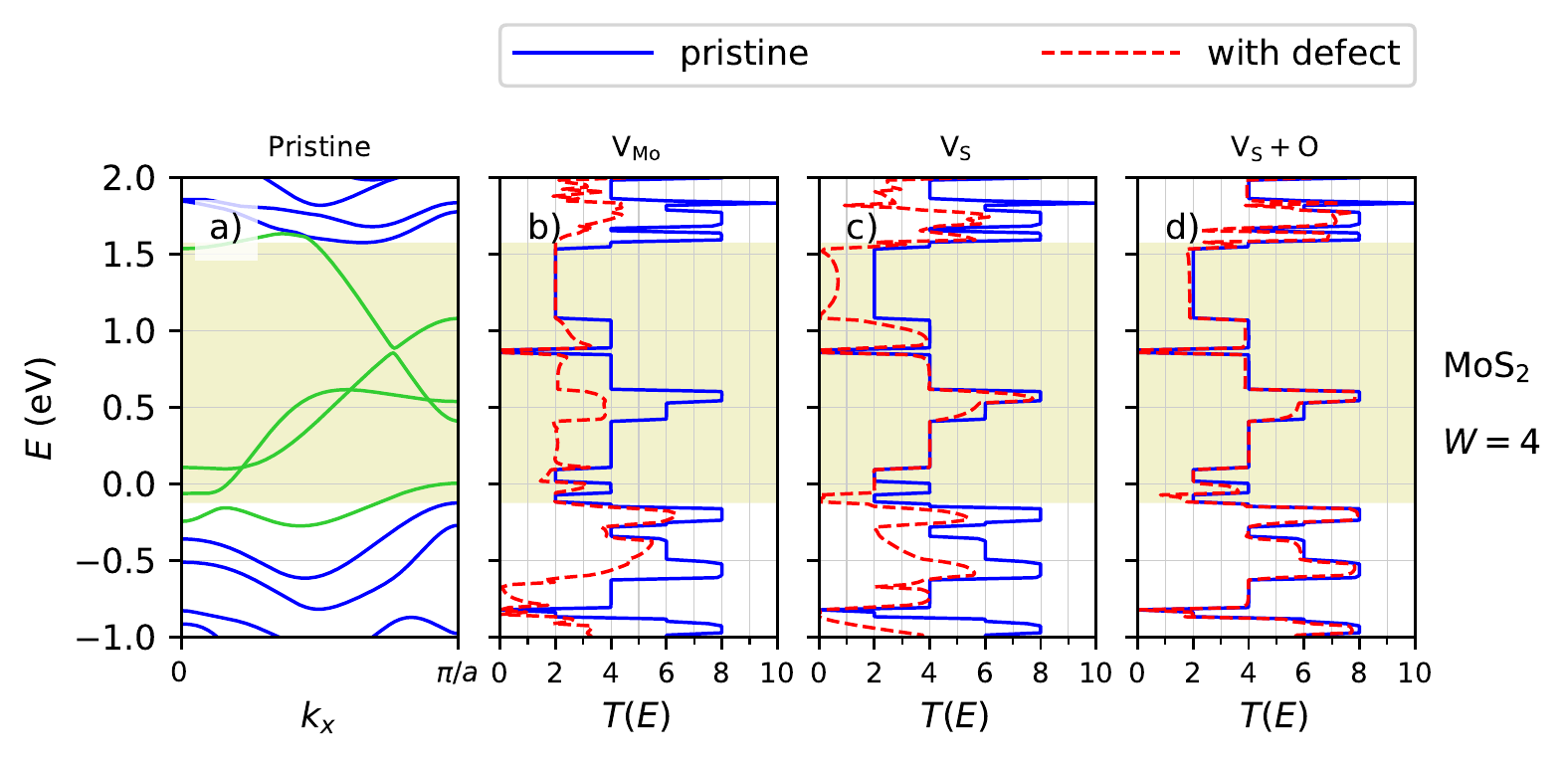}
	\caption{Transmission spectrum (TS) of MoS2 zigzag nanoribbons of width $W=4$ in presence of one of the following edge defects: Mo vacancy (b); S vacancy (c); Oxygen-saturated S vacancy (d). The TS for a pristine ribbon is also shown for comparison in each panel, and its band structure is shown in panel a. The energy region of insulating bulk is highlighted in yellow.}
	\label{fig:TS_edge-zig_4-MoS2}
\end{figure}

In Fig.\ \ref{fig:TS_edge-zig_4-MoS2} we report the transmission spectrum for an MoS$_2$ zigzag nanoribbon of width $W=4$. As for the case of $W=8$, both Mo and S vacancies lead to a very strong suppression of the transmission spectrum, while chemical saturation with Oxygen restores the transport properties of pristine nanoribbons, although not perfectly. 

\FloatBarrier
\section{Disordered configurations}

\begin{figure}
	\centering
	\includegraphics[width=\linewidth]{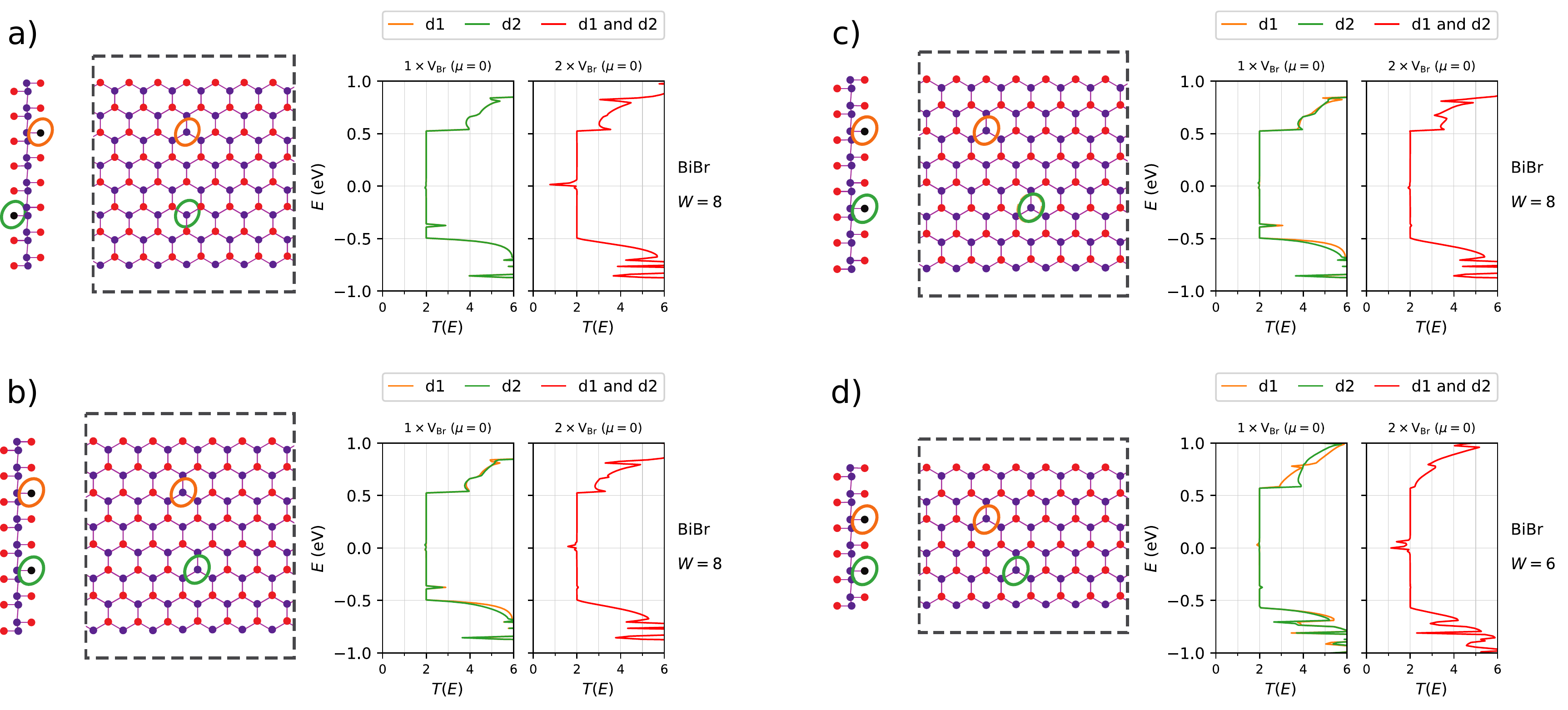}
	\caption{Transmission spectrum (TS) of BiBr zigzag nanoribbons of different widths in presence of multiple non-magnetic bulk defects. Positions of the impurities (denoted d1 and d2) are reported in the corresponding structures. In all cases, we compare the TS in the presence of d1 only, d2 only, and d1 and d2 simultaneously.}	
	\label{fig:disorder}
\end{figure}

In this section we compare the transmission function of different nanoribbon configurations with a pair of non-magnetic bulk defects. 
The structure in Fig.\ \ref{fig:disorder}a corresponds to the one discussed in the main text, while panels b, c, and d pertain to different disordered configurations and nanoribbon widths.
In all cases a single bulk impurity is not sufficient to generate inter-edge backscattering. Panels a, b and d show instead that multiple impurities generate a dip in the transmission function around $E \approx 0$ due to inter-edge coupling.
In panel c, the two Br vacancies appear to be too far from each other to generate backscattering.

\FloatBarrier
\section{Local density of states}

\begin{figure}
	\centering
	\includegraphics[width=\linewidth]{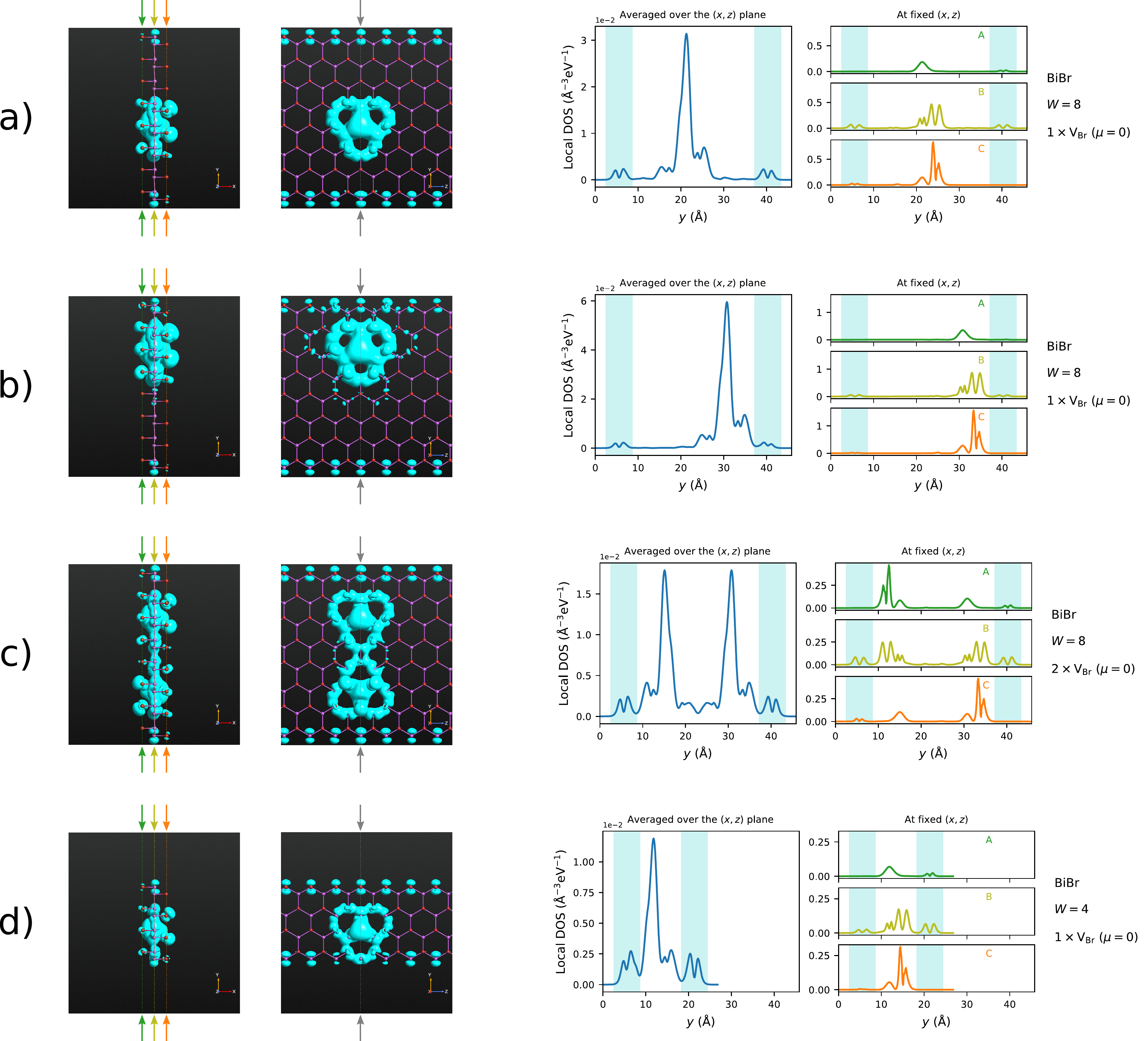}
	\caption{Local density of states (LDOS) for different nanoribbon widths and disordered configurations according to the legend. For each panel, the first and second columns show side and top views of the surface $\mathrm{LDOS}(x,y,z) = 0.02 \,\mathrm{\AA}^{-3} \mathrm{eV}^{-1}$ respectively. The third column corresponds to the LDOS in the transverse direction $y$ averaged over the $xz$ plane, while the last column shows the LDOS along three different lines as reported on the left. The spatial region spanned by the edge states is highlighted in cyan in the last two cases.}	
	\label{fig:LDOS}
\end{figure}

To compare the spatial extension of edge and defect states, we study here the local density of states (LDOS) for different device configurations.
As demonstrated by Fig.\ \ref{fig:LDOS}a and \ref{fig:LDOS}b, a single bulk defect state cannot overlap significantly with both edge states in a wide nanoribbon, so that impurity-mediated backscattering through one single vacancy is not permitted.
On the other hand, a significant overlap is present in the cases of multiple impurities (panel c) or narrow structures (panel d).

\FloatBarrier
\section{Magnetic moments of edge and bulk impurities}
\label{subsec:add}

\begin{figure}
	\centering
	\includegraphics[width=0.5\linewidth]{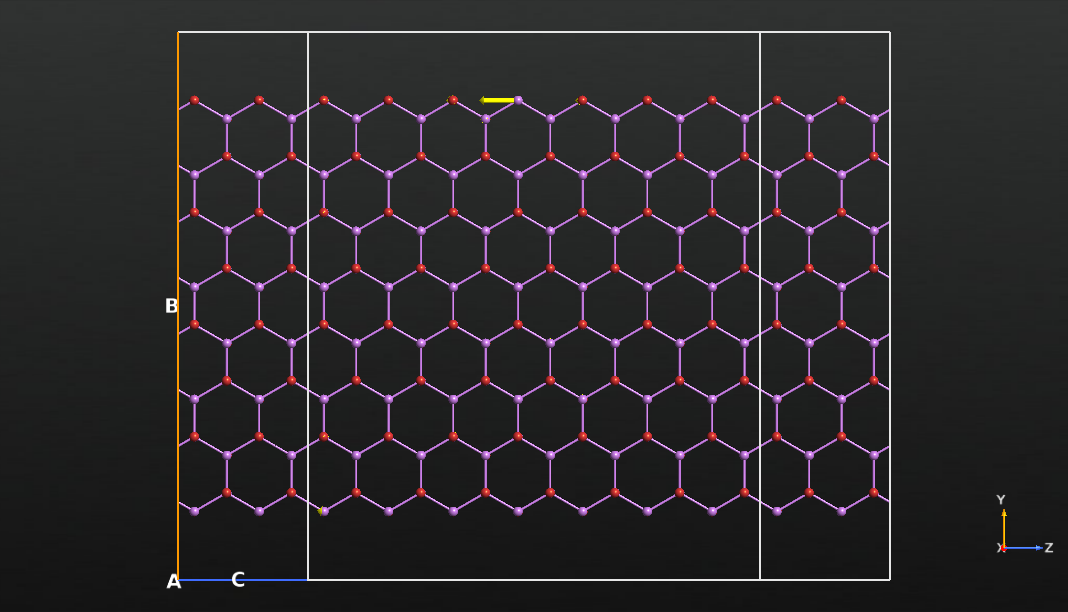}
	\caption{Magnetic moments of an SbBr zigzag nanoribbon of width $W=8$ calculated from a Mulliken population analysis. Sb atoms are in purple, while Br atoms are in red. A yellow arrow denotes the local spin direction and the magnitude of the magnetic moment.}
	\label{fig:mag_mom}
\end{figure}

Here we report the magnetic moments of all the magnetic device configurations we have found.

We have extracted the magnetic moments from a Mulliken population analysis by subtracting the number of spin down electrons from the number of spin up electrons at each site.
We interpret the number obtained in this way as the local magnetic moment in units of the Bohr magneton $\mu_\mathrm{B}$.
Note that the Mulliken population for this particular case of non-collinear spin calculation is a four component spin tensor, which is separately diagonalized at each site to give a local spin direction.

Generally, magnetic moments are zero everywhere except in the immediate vicinity of the defect, as shown in Figure \ref{fig:mag_mom}. Therefore, we only show the magnitude of the magnetic moment at the Bimsuth (Antimony) site closest to the vacancy. The corresponding results are given in Table \ref{tab:magnetic_moments}.

\begin{table}
	\centering
	\begin{tabular}{lccc}
		material & W & $\mu$ edge defect ($\mu_\mathrm{B}$) & $\mu$ bulk defect ($\mu_\mathrm{B}$)\\
		\hline
		BiF &	4 &	0.015 &	0.042\\
		BiF &	8 &	0.017 &	0.072\\
		\hline
		BiCl & 	4 &	0.018 &	0.370\\
		BiCl & 	8 &	0.018 &	0.401\\
		\hline
		BiBr &	4 &	0.001 &	0.298\\
		BiBr &	8 &	0.000 &	0.361\\
		\hline
		BiI &	4 & 0.010 &	0.289\\
		BiI &	8 & 0.015 &	0.319\\
		\hline
		SbF & 	4 & 0.658 &	0.491\\
		SbF & 	8 & 0.651 &	0.539\\		
		\hline
		SbCl & 	4 &	0.960 &	0.710\\
		SbCl & 	8 &	0.945 &	0.661\\
		\hline
		SbBr &	4 &	0.933 &	0.752\\
		SbBr &	8 &	0.916 &	0.702\\
		\hline
		SbI &	4 &	0.860 &	0.689\\
		SbI &	8 &	0.857 &	0.627\\
		\hline
	\end{tabular}	
	\caption{Magnetic moments calculated at the Bi (Sb) sites closest to the vacancy defect. Two different nanoribbon widths are considered here ($W=4$ and $W=8$).}
	\label{tab:magnetic_moments}
\end{table}

We note that Antimony-based structures always show a sizable magnetic moment of the order of $0.5-0.9 \mu_\mathrm{B}$. On the other hand, Bismuth-based structures behave quite differently depending on the position of the impurity. They are almost non-magnetic in the edge defect configuration, with $\mu \approx 0.01 \mu_\mathrm{B}$, while they bear a magnetic moment $\mu = 0.3-0.4 \mu_\mathrm{B}$ in the bulk defect configuration --- the only exception being a bulk defect in BiF, with a much smaller moment $\mu = 0.04-0.07 \mu_\mathrm{B}$.

\bibliography{biblio.bib}